\documentclass[3p]{elsarticle}

\usepackage{amsmath}
\usepackage{amsthm}
\usepackage{amssymb}
\usepackage{graphicx}
\usepackage{esint}
\usepackage{xcolor}

\usepackage{lineno}
\PassOptionsToPackage{hyphens}{url}
\usepackage[pdfencoding=auto,psdextra]{hyperref}
\usepackage{bookmark}

\biboptions{numbers,sort&compress}

\journal{Physica E: Low-Dimensional Systems and Nanostructures}

\begin{document}

\begin{frontmatter}

\title{Launching of Davydov solitons in protein $\alpha$-helix spines}

\author[address1]{Danko D. Georgiev\corref{mycorrespondingauthor}}
\ead{danko.georgiev@mail.bg}
\cortext[mycorrespondingauthor]{Corresponding author}

\author[address2]{James F. Glazebrook}
\ead{jfglazebrook@eiu.edu}

\address[address1]{Institute for Advanced Study, 30 Vasilaki Papadopulu Str., Varna 9010, Bulgaria}
\address[address2]{Department of Mathematics and Computer Science, Eastern Illinois University, Charleston, IL 61920, USA}

\begin{abstract}
The biological order provided by $\alpha$-helical secondary protein structures is an important resource exploitable by living organisms for increasing the efficiency of energy transport. In particular, self-trapping of amide~I energy quanta by the induced phonon deformation of the hydrogen-bonded lattice of peptide groups is capable of generating either pinned or moving solitary waves following the Davydov quasiparticle/soliton model. The effect of applied in-phase Gaussian pulses of amide~I energy, however, was found to be strongly dependent on the site of application. Moving solitons were only launched when the amide~I energy was applied at one of the $\alpha$-helix ends, whereas pinned solitons were produced in the $\alpha$-helix interior. In this paper, we describe a general mechanism that launches moving solitons in the interior of the $\alpha$-helix through phase-modulated Gaussian pulses of amide~I energy. We also compare the predicted soliton velocity based on effective soliton mass and the observed soliton velocity in computer simulations for different parameter values of the isotropy of the exciton--phonon interaction. The presented results demonstrate the capacity for explicit control of the soliton velocity in protein $\alpha$-helices, and further support the plausibility of gradual optimization of quantum dynamics for achieving specialized protein functions through natural selection.
\end{abstract}


\date{June 26, 2020}

\end{frontmatter}


\section{Introduction}

Proteins play a diverse number of roles in living organisms. In the impressive portfolio of protein assignments, the most important are the following: catalytic, structural, contractile, regulatory, protective, and transportation. The biological value of proteins is paramount for organisms when passing the former to their offspring, in the form of hereditary DNA information, to the extent that exact amino acid sequences of the proteins have already ensured the success of the parent organism. Each of the listed protein functions is performed at the expense of free biochemical energy and is supported by highly evolved specialized protein structural motifs. While the dynamics of all biomolecules is fundamentally governed by quantum mechanical laws, the highly efficient utilization of individual energy quanta by proteins suggests that characteristic quantum effects may be indispensable for a deeper understanding of the mechanisms of life \cite{GeorgievGlazebrook2020}.

Protein $\alpha$-helices constitute a common motif in the secondary
structure of proteins \cite{Scholtz1992,Kohn1997,Haimov2016}. Each
$\alpha$-helix is a right-handed spiral with 3.6 amino acid residues
per turn \cite{Pauling1951a,GeorgievGlazebrook2019a}. The helical structure is stabilized
by hydrogen bonding between the N--H group from an amino acid and
the C=O group from another amino acid located four residues earlier
in the primary amino acid sequence of the protein. This bonding produces
three parallel chains of hydrogen-bonded peptide groups referred to
as $\alpha$-helix spines \cite{GeorgievGlazebrook2019a}. The interaction
between C=O bond stretching (amide~I excitons) and the deformation
of the lattice of --C=O$\cdots$H--N-- hydrogen bonds (phonons) provides
a quantum model for the transport of energy in terms of Davydov solitons
\cite{Davydov1973,Davydov1976,Davydov1979,Davydov1981,Davydov1982,Davydov1986,Davydov1986b,Davydov1987,Davydov1988,Kivshar1989}.

For a single $\alpha$-helix spine of hydrogen-bonded peptide groups,
the generalized Davydov model describes the quantum dynamics of amide~I excitons, lattice phonons, and the nonlinear exciton--phonon interaction \cite{Brizhik1983,Brizhik1988,Brizhik1993,Brizhik1995,Brizhik2004,Brizhik2006,Brizhik2010,Cruzeiro1988,Cruzeiro1994,Cruzeiro1997,Cruzeiro2009,Kerr1987,Kerr1990,MacNeil1984,Scott1984,Scott1985,Scott1992,Luo2011,Luo2017,Sun2010}.
The system of gauge transformed quantum equations of motion is \cite{GeorgievGlazebrook2020}
\begin{eqnarray}
\imath\hbar\frac{d}{dt}a_{n} & = & -J_{n+1}a_{n+1}-J_{n}a_{n-1}+\chi\left[b_{n+1}+(\xi-1)b_{n}-\xi b_{n-1}\right]a_{n}\label{eq:gauge-1}\\
M_{n}\frac{d^{2}}{dt^{2}}b_{n} & = & w\Big(b_{n-1}-2b_{n}+b_{n+1})-Q\chi\Big(\left|a_{n-1}\right|^{2}+(\xi-1)\left|a_{n}\right|^{2}-\xi\left|a_{n+1}\right|^{2}\Big)\label{eq:gauge-2}
\end{eqnarray}
where $Q$ is the number of excited amide~I quanta, $a_n$ is the amide~I quantum probability amplitude at the $n$th peptide group, $\hbar$ is the reduced Planck constant, $J_n$~is the dipole--dipole coupling energy between the $n$th and $(n-1)$th amide~I oscillator along
the $\alpha$-helix spine, $b_n$ is the phonon displacement of the
$n$th peptide group from its equilibrium position, $M_n$ is the mass of the $n$th peptide group,
$w$ is the spring constant of the hydrogen bonds in the lattice \cite{Davydov1976,Davydov1979,Scott1992},
$\chi=\bar{\chi}\frac{2}{1+\xi}$ is a nonlinear coupling parameter between the amide~I exciton and the phonon displacements of peptide groups in the lattice of hydrogen bonds, $\bar{\chi}=\frac{\chi_{r}+\chi_{l}}{2}$, $\chi_{r}$~and~$\chi_{l}$ couple the amide~I oscillator (C=O~group) with the hydrogen bonds ($\cdots$) respectively to the right or to the left in the chemical structure of the $\alpha$-helix spine
\begin{equation*}
\cdots H-N-C=O\underset{\chi_{l}}{\underbrace{\cdots H-N-C=}}\underset{\chi_{r}}{\underbrace{O\cdots H}}-N-C=O\cdots
\end{equation*}
and $\xi=\frac{\chi_{l}}{\chi_{r}}$ is the anisotropy parameter of
the exciton--phonon interaction (in proteins $\chi_{r}> 0$ and
\mbox{$0\leq\chi_{l}\leq\chi_{r}$} thereby constraining $\xi$ in the interval
$[0,1]$) \cite{Luo2017,GeorgievGlazebrook2019a,GeorgievGlazebrook2019b}. For a complete derivation of the system of quantum equations of motion, we refer the interested reader to Ref.~\cite{GeorgievGlazebrook2020}.

Most biochemical processes in living organisms are driven by free energy as
provided by the cleavage of pyrophosphate bonds in adenosine triphosphate~(ATP). ATP hydrolysis to adenosine diphosphate (ADP) and orthophosphate (P$_\textrm{i}$) releases
$0.41$~eV of free energy. Thus, for every ATP molecule utilized,
proteins are able to excite $Q=2$ amide~I quanta, each of which carries $0.205$~eV.

In order to study the mechanisms for launching of Davydov solitons, which represent traveling solitary waves of amide~I energy dressed by an accompanying phonon deformation in the protein $\alpha$-helix, we have integrated numerically the system of Davydov equations \eqref{eq:gauge-1} and \eqref{eq:gauge-2} using uniform values $J$ and $M$ for all peptide groups $n$. For enhancement of the accuracy and speed of computational simulations, we performed the numerical integration with LSODA, Livermore Solver for Ordinary Differential equations with Automatic selection between nonstiff (Adams) and stiff (Backward Differentiation Formula, BDF) methods \cite{Petzold1983,Hindmarsh1983,Hindmarsh1995,Trott2006,GeorgievGlazebrook2019b}.

The main problems that we address in this present computational study are: Is it possible to generate Davydov solitons without an initial phonon dressing? Is there a physical mechanism that would allow proteins to launch Davydov solitons in the interior of protein $\alpha$-helices? Is the higher flux of amide I energy quanta capable of supporting Davydov solitons in protein conformations with weak nonlinear exciton--phonon coupling? Is the analytic continuum approximation adequate for predicting the properties of Davydov solitons in the case of anisotropic exciton-phonon interaction? The positive answers to the first three questions support the plausibility of Davydov solitons as a physical mechanism employed in the evolution of protein function through gradual steps favored by natural selection. The partially negative answer to the last question highlights the necessity of combining both computational and analytic approaches for studying the effects of lattice discreteness on the quantum transport of energy in proteins.

\section{Computational study}

\subsection{\label{sub:Model}Model parameters}

The solitons described here are supported by the system of generalized Davydov equations
\eqref{eq:gauge-1} and \eqref{eq:gauge-2} for a wide range of the
biophysical parameters characterizing the protein $\alpha$-helix
\cite{Brizhik1983,Brizhik1988,Brizhik1993,Brizhik1995,Brizhik2004,Brizhik2006,Brizhik2010,Brizhik2015,MacNeil1984,Scott1984,Scott1985,Cruzeiro1988,GeorgievGlazebrook2019a,GeorgievGlazebrook2019b}.
In order to compare the results in this present work with
those reported in previous studies, we have fixed the model parameters
as follows: spring constant of the hydrogen bonds in the lattice $w=13$~N/m \cite{Itoh1972}, nonlinear exciton--phonon coupling parameter $\bar{\chi}=35$~pN \cite{GeorgievGlazebrook2019a},
uniform mass of amino acid residues $M=1.9\times10^{-25}$~kg \cite{Cruzeiro1988}, and uniform amide~I dipole--dipole coupling energy $J=0.155$~zJ \cite{Nevskaya1976}.

The solitons were produced in the absence of initial phonon dressing of
the applied amide~I energy pulses. Thus, the initial state at $t=0$
of the lattice of hydrogen bonds was considered to be unperturbed,
$b_{n}(0)=0$ and $\frac{d}{dt}b_{n}(0)=0$. Therefore, all of the
energy supplied to the protein $\alpha$-helix is due to a fixed non-zero
initial exciton distribution $a_{n}(0)$ with $Q\geq1$. To furnish
a sufficient arena for phase modulation through phase parameter $\Delta\omega$,
individual Gaussian pulses of amide~I energy were spread over 7 peptide
groups with quantum probability amplitudes $a_{n}$ given by
\begin{equation}
\{A_{3}e^{-\imath3\Delta\omega},A_{2}e^{-\imath2\Delta\omega},A_{1}e^{-\imath1\Delta\omega},A_{0}e^{\imath0\Delta\omega},A_{1}e^{+\imath1\Delta\omega},A_{2}e^{+\imath2\Delta\omega},A_{3}e^{+\imath3\Delta\omega}\}\label{eq:Gaussian-7}
\end{equation}
where $A_{0}=\sqrt{0.286}$, $A_{1}=\sqrt{0.222}$, $A_{2}=\sqrt{0.105}$
and $A_{3}=\sqrt{0.03}$.

Denoting the longitudinal axis of the $\alpha$-helix spine by $x$, we can
express the initial discrete distribution of exciton quantum probability
amplitudes centered at $x_0$ as
\begin{equation}
a(x)=\sum_{j=-3}^{3}\frac{1}{\sqrt{r}}A_{\left|j\right|}e^{\imath j\Delta\omega}\left[\Theta\left(\frac{x-x_0}{r}-j+\frac{1}{2}\right)-\Theta\left(\frac{x-x_0}{r}-j-\frac{1}{2}\right)\right]
\end{equation}
where $\Theta\left(x\right)=\frac{d}{dx}\max\left\{ x,0\right\} $
is the Heaviside step function and $r=0.45$ nm is the distance between
neighboring peptide groups along the spine. This initial exciton state
is normalized
\begin{equation}
\int_{-\infty}^{\infty}\left|a(x)\right|^{2} dx=A_{0}^{2}+2\left(A_{1}^{2}+A_{2}^{2}+A_{3}^{2}\right)=1.
\end{equation}
Application of a Fourier transform determines the exciton wavefunction in wavenumber basis~$k$ as
\begin{equation}
a(k)=\sqrt{\frac{2}{k^{2}\pi r}}\sin\left(\frac{kr}{2}\right)e^{-\imath k x_0}\sum_{j=-3}^{3}A_{\left|j\right|}e^{\imath j\left(kr-\Delta\omega\right)} .
\end{equation}
The wavenumber quantum probability distribution is
\begin{equation}
\left|a(k)\right|^{2}=\frac{8\sin^{2}\left(\frac{kr}{2}\right)}{k^{2}\pi r}\left\{ \frac{A_{0}}{2}+\sum_{j=1}^{3}A_{j}\cos\left[j\left(kr-\Delta\omega\right)\right]\right\} ^{2}
\end{equation}
with corresponding expectation value
\begin{equation}
\langle\hat{k}\rangle=\int_{-\infty}^{\infty}\left|a(k)\right|^{2}k\,dk=\frac{2\sin\left(\Delta\omega\right)}{r}\left(A_{0}A_{1}+A_{1}A_{2}+A_{2}A_{3}\right) . \label{eq:ave-k}
\end{equation}

Due to the initial unperturbed state of the phonon lattice, there
is no phonon dressing of the exciton at $t=0$, namely $\langle\Psi(0)|\hat{H}_{\textrm{ph}}|\Psi(0)\rangle=0$
and $\langle\Psi(0)|\hat{H}_{\textrm{int}}|\Psi(0)\rangle=0$, where the phonon Hamiltonian $\hat{H}_\textrm{ph}$ and the exciton-phonon interaction Hamiltonian $\hat{H}_\textrm{int}$ are given by \cite{GeorgievGlazebrook2020}
\begin{eqnarray}
\hat{H}_\textrm{ph} &=& \frac{1}{2}\sum_{n}\left[\frac{\hat{p}_{n}^{2}}{M_{n}}+w\left(\hat{u}_{n+1}-\hat{u}_{n}\right)^{2}\right],\\
\hat{H}_\textrm{int} &=& \chi\sum_{n}\left(\hat{u}_{n+1}+\left(\xi-1\right)\hat{u}_{n}-\xi\hat{u}_{n-1}\right)\hat{a}_{n}^{\dagger}\hat{a}_{n},
\end{eqnarray}
$\hat{a}_n^{\dagger}$ and $\hat{a}_n$ are the boson creation and annihilation operators for the amide~I excitons, $\hat{p}_n$~is the momentum operator and $\hat{u}_n$
is the displacement operator from the equilibrium position of the $n$th
peptide group, and $|\Psi(0)\rangle$ is the initial state at $t=0$ of the generalized ansatz state vector \cite{Kerr1990}
\begin{equation}
|\Psi(t)\rangle=|\psi_\textrm{ex}(t)\rangle|\psi_\textrm{ph}(t)\rangle\label{eq:ansatz}
\end{equation}
comprised of a Hartree approximate eigenstate for the excitons \cite{Scott1992,Lai1989a,Lai1989b,Wright1990}
\begin{equation}
|\psi_\textrm{ex}(t)\rangle=\frac{1}{\sqrt{Q!}}\left[\sum_{n}a_{n}(t)\hat{a}_{n}^{\dagger}\right]^{Q}|0_{\textrm{ex}}\rangle
\end{equation}
and a Glauber coherent phonon state for the lattice \cite{Scott1992,Glauber1963a,Glauber1963b}
\begin{equation}
|\psi_\textrm{ph}(t)\rangle=e^{-\frac{\imath}{\hbar}\sum_{j}\left(b_{j}(t)\hat{p}_{j}-c_{j}(t)\hat{u}_{j}\right)}|0_{\textrm{ph}}\rangle .
\end{equation}

Because the total energy of the composite system is conserved during temporal
evolution according to the Schr\"{o}dinger equation, the soliton energy
could be calculated from the initial expectation value of the exciton
Hamiltonian
\begin{equation}
\hat{H}_\textrm{ex} = \sum_{n}\left[-J_{n+1}\hat{a}_{n}^{\dagger}\hat{a}_{n+1}-J_{n}\hat{a}_{n}^{\dagger}\hat{a}_{n-1}\right] .
\end{equation}
After setting all dipole--dipole coupling energies to be the
same $J_{n}=J$, we obtain
\begin{align}
\langle\Psi(0)|\hat{H}_{\textrm{ex}}|\Psi(0)\rangle & =-QJ\sum_{n}\left[a_{n}^{*}(0)a_{n+1}(0)+a_{n}(0)a_{n+1}^{*}(0)\right]\nonumber \\
 & =-4QJ\cos\left(\Delta\omega\right)\left(A_{0}A_{1}+A_{1}A_{2}+A_{2}A_{3}\right) . \label{eq:ex-energy}
\end{align}

In the simulations with two amide~I energy quanta ($Q=2$), the exciton
probability amplitudes of individual solitons \eqref{eq:Gaussian-7}
were multiplied by a factor of $\sqrt{\frac{1}{Q}}$ for two colliding
single solitons or by a factor of $\sqrt{\frac{2}{Q}}$ for a single
double soliton.

\subsection{Effect of phase-modulation on single solitons}

Extensive previous research with reflective boundary conditions has
shown that in-phase Gaussian pulses of amide~I energy generate moving
Davydov solitons only when applied at one of the two ends of a protein
$\alpha$-helix, whereas the same pulses produce pinned solitons in
the interior of the $\alpha$-helix \cite{GeorgievGlazebrook2019a,GeorgievGlazebrook2019b,GeorgievGlazebrook2020}.
Because the biological function of proteins in living matter may require
delivery and transport of energy irrespective of the application site,
we have explored the possibility of utilizing phase-modulation to
launch moving solitons in the interior of protein $\alpha$-helices.

To investigate numerically how phase-modulation $\Delta\omega\in[-\pi,\pi]$
affects the velocity of generated Davydov solitons, we have applied
Gaussian pulses of amide~I energy centered at the peptide group $n=10$
of a protein $\alpha$-helix spine comprised of $n_{\max}=40$ peptide
groups. For single solitons, a single quantum of amide~I energy ($Q=1$)
was used.

\begin{figure}[t]
\begin{centering}
\includegraphics[width=135mm]{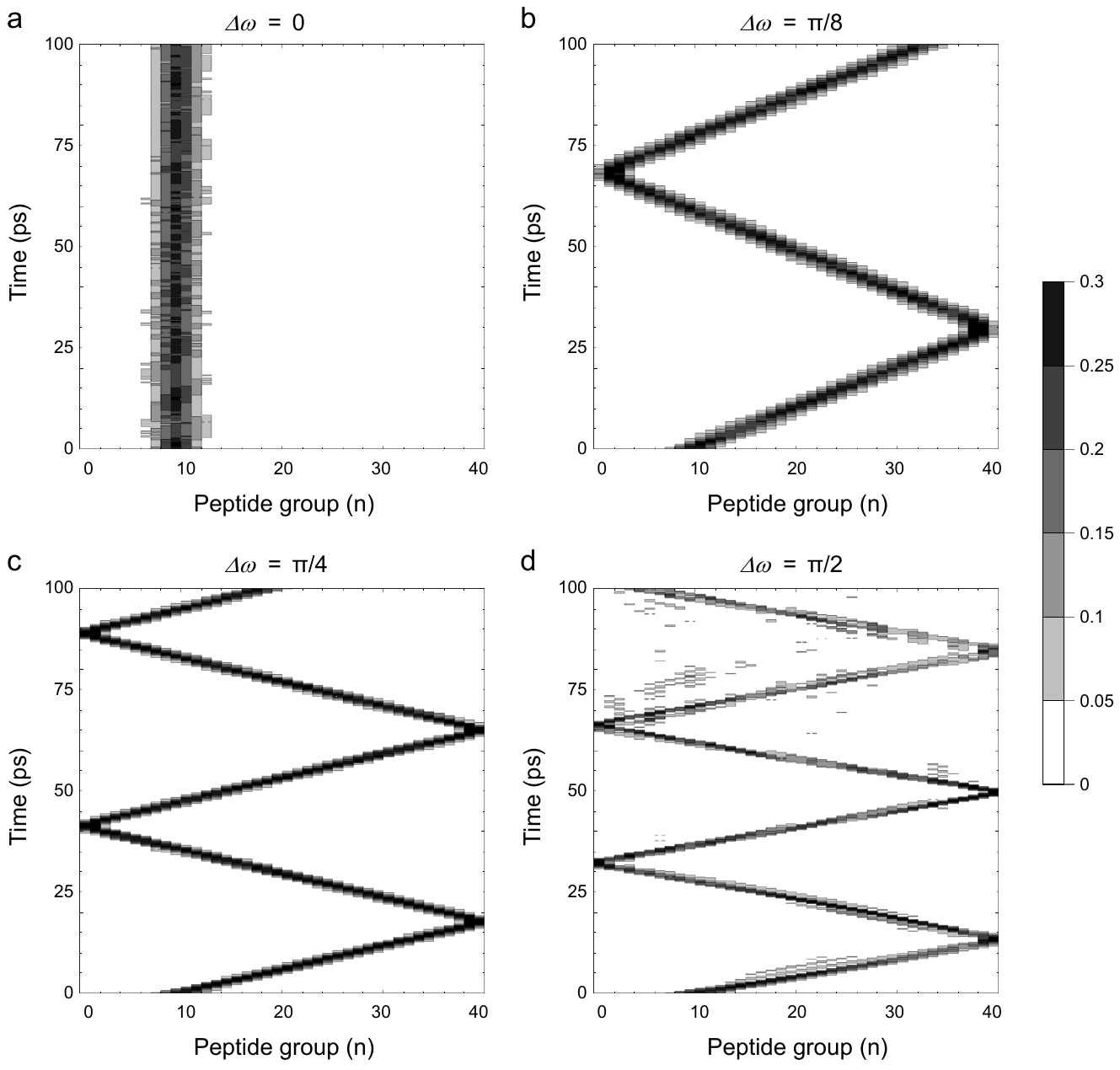}
\par\end{centering}

\caption{\label{fig:1}Single soliton dynamics for $Q=1$ and $\xi=1$, visualized in a contour plot of the amide~I exciton expectation value $Q|a_{n}|^{2}$. The applied energy pulse is centered at the peptide group $n=10$ of a protein $\alpha$-helix spine comprised of $n_{\max}=40$ peptide
groups. In the absence of phase modulation $\Delta\omega=0$ (a),
the soliton is pinned. In the presence of positive phase modulation,
$\Delta\omega=\frac{\pi}{8}$ (b), $\Delta\omega=\frac{\pi}{4}$ (c)
or $\Delta\omega=\frac{\pi}{2}$ (d), the soliton moves to the right.}
\end{figure}

\begin{figure}[t]
\begin{centering}
\includegraphics[width=135mm]{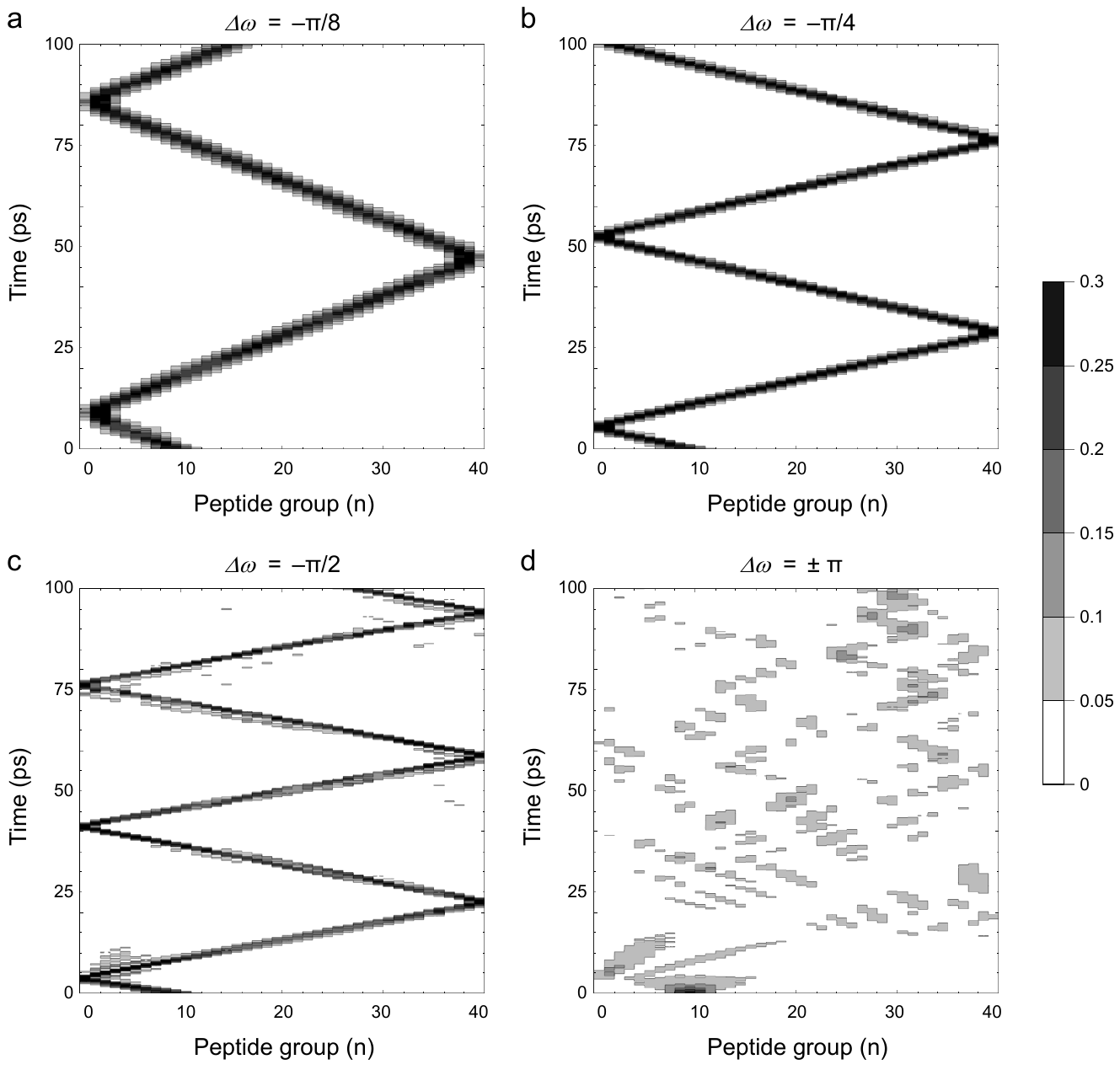}
\par\end{centering}

\caption{\label{fig:2}Single soliton dynamics for $Q=1$ and $\xi=1$, visualized in a contour plot of the amide~I exciton expectation value $Q|a_{n}|^{2}$. The applied energy pulse is centered at the peptide group $n=10$ of a protein $\alpha$-helix spine comprised of $n_{\max}=40$ peptide
groups. In the presence of negative phase modulation, $\Delta\omega=-\frac{\pi}{8}$
(a), $\Delta\omega=-\frac{\pi}{4}$ (b) or $\Delta\omega=-\frac{\pi}{2}$
(c), the soliton moves to the left. At $\Delta\omega=\pm\pi$ (d),
the soliton is unstable and quickly disintegrates.}
\end{figure}

For completely isotropic exciton--phonon interaction $\xi=1$, in
the absence of phase modulation ($\Delta\omega=0$) the generated
soliton remained pinned at the place of origin (Fig.~\ref{fig:1}a,
Video~1). In the presence of positive phase modulation $(0<\Delta\omega\leq\frac{\pi}{2})$,
the soliton moves to the right (Fig.~\ref{fig:1}b-d, Videos~2-3),
whereas for negative phase modulation $(-\frac{\pi}{2}\leq\Delta\omega<0)$,
the soliton moves to the left (Fig.~\ref{fig:2}a-c). At $\Delta\omega=\pm\pi$,
the soliton appears to be unstable and quickly disintegrates (Fig.~\ref{fig:2}d,
Video~4).

\begin{figure}[t]
\begin{centering}
\includegraphics[width=135mm]{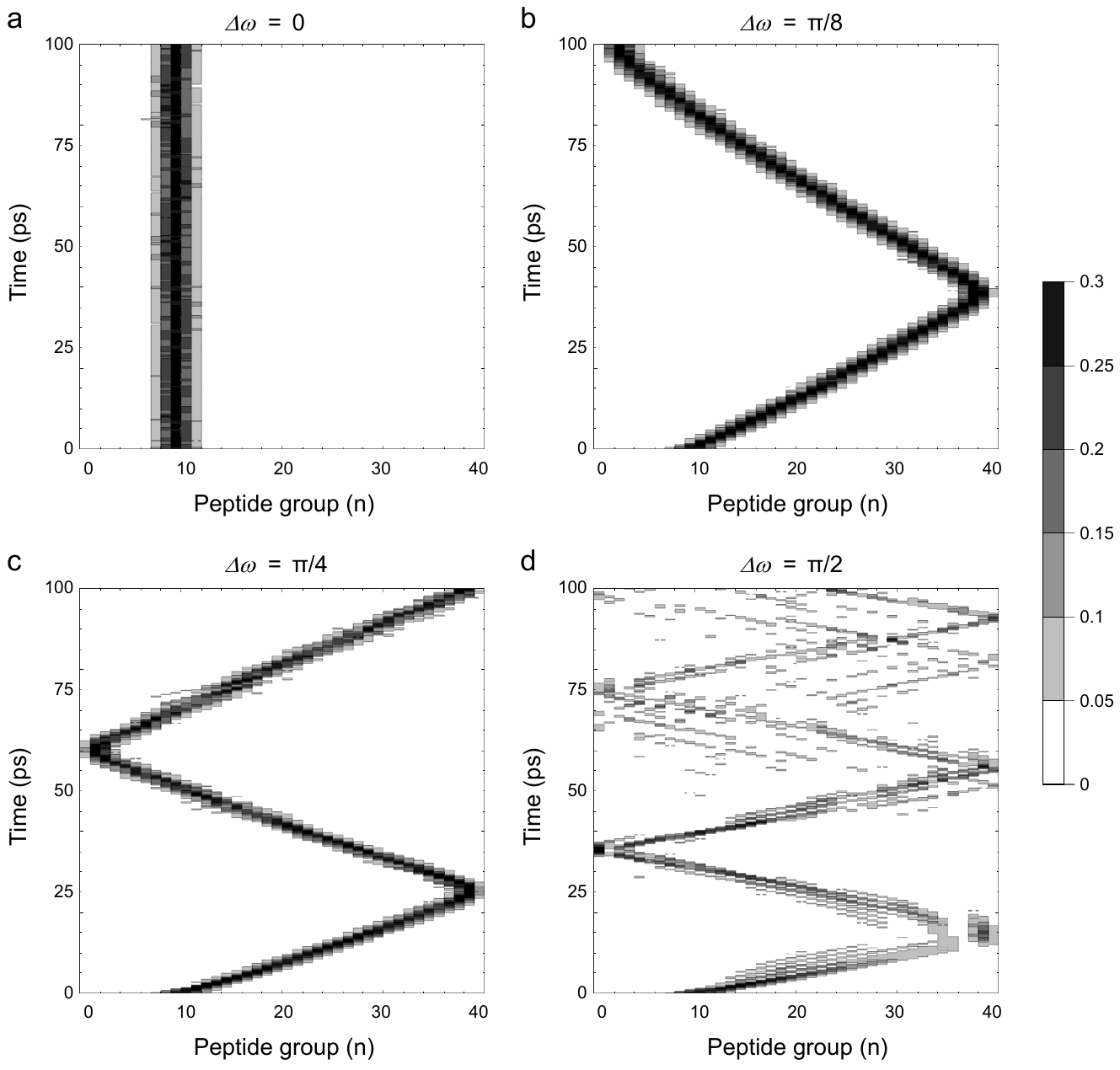}
\par\end{centering}

\caption{\label{fig:3}Single soliton dynamics for $Q=1$ and $\xi=0$, visualized in a contour plot of the amide~I exciton expectation value $Q|a_{n}|^{2}$. The applied energy pulse is centered at the peptide group $n=10$ of a protein $\alpha$-helix spine comprised of $n_{\max}=40$ peptide
groups. In the absence of phase modulation $\Delta\omega=0$ (a),
the soliton is pinned. In the presence of positive phase modulation,
$\Delta\omega=\frac{\pi}{8}$ (b) or $\Delta\omega=\frac{\pi}{4}$
(c) the soliton moves to the right. For $\Delta\omega=\frac{\pi}{2}$
(d), the soliton is unstable.}
\end{figure}

\begin{figure}[t]
\begin{centering}
\includegraphics[width=135mm]{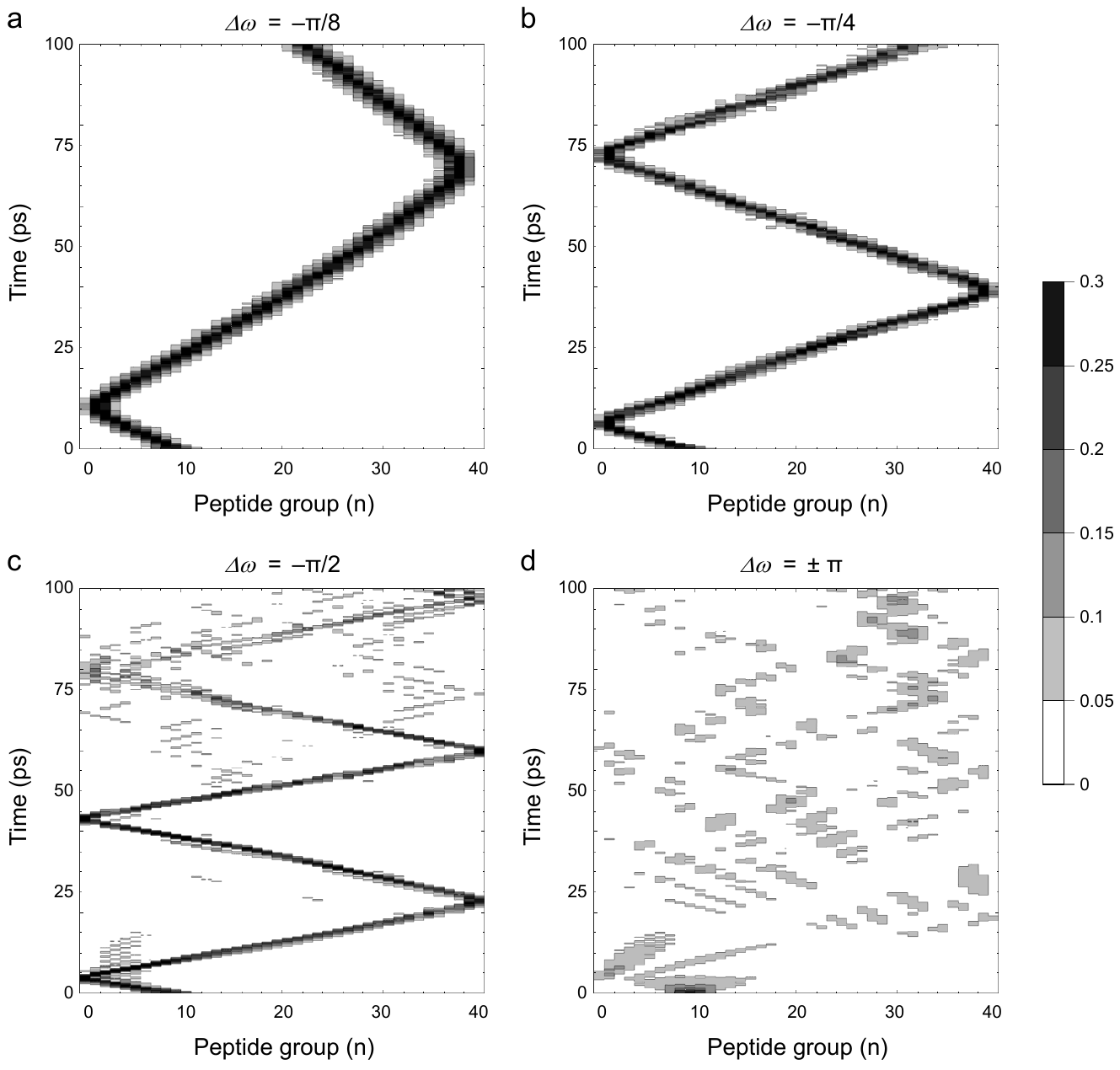}
\par\end{centering}

\caption{\label{fig:4}Single soliton dynamics for $Q=1$ and $\xi=0$, visualized in a contour plot of the amide~I exciton expectation value $Q|a_{n}|^{2}$. The applied energy pulse is centered at the peptide group $n=10$ of a protein $\alpha$-helix spine comprised of $n_{\max}=40$ peptide
groups. In the presence of negative phase modulation, $\Delta\omega=-\frac{\pi}{8}$
(a) or $\Delta\omega=-\frac{\pi}{4}$ (b), the soliton moves to the
left. For $\Delta\omega=-\frac{\pi}{2}$ (c) the soliton is unstable,
and at $\Delta\omega=\pm\pi$ (d) it quickly disintegrates.}
\end{figure}

The results of the simulations were qualitatively similar to those for completely anisotropic
exciton--phonon interaction $\xi=0$. In the absence of phase modulation
($\Delta\omega=0$), the generated soliton remained pinned at the place
of origin (Fig.~\ref{fig:3}a). In the presence of positive phase
modulation $(0<\Delta\omega<\frac{\pi}{2})$, the soliton moves to
the right (Fig.~\ref{fig:3}b-c), whereas for negative phase modulation
$(-\frac{\pi}{2}<\Delta\omega<0)$, the soliton moves to the left
(Fig.~\ref{fig:4}a-b). Crucially, in this case the soliton instability
is already pronounced at $\Delta\omega=\pm\frac{\pi}{2}$ (Figs.~\ref{fig:3}d
and \ref{fig:4}c), and the soliton also quickly disintegrates at
$\Delta\omega=\pm\pi$ (Fig.~\ref{fig:4}d).

Because the total energy of the composite system is conserved throughout
the simulations, it is equal to the initial exciton energy \eqref{eq:ex-energy}
applied to the protein $\alpha$-helix. For $\Delta\omega\in[0,\pm\frac{\pi}{2})$
the total energy stays negative, which is consistent with the observed
persistence of the Davydov solitons as launched. However, when the phase
modulation is $\Delta\omega\in(\pm\frac{\pi}{2},\pm\pi]$, the total
energy becomes positive and this facilitates dispersal of the exciton
quantum probability amplitudes without any soliton formation.

Previous analytical models \cite{Davydov1979,Davydov1981,Davydov1986b,Scott1992}
of an $\alpha$-helix spine have determined the effective mass of
a single exciton
\begin{equation}
m_{\textrm{ex}}=\frac{\hbar^{2}}{2r^{2}J}
\end{equation}
and the effective mass of a single Davydov soliton \cite{Davydov1986b}
\begin{equation}
m_{\textrm{sol}}=m_{\textrm{ex}}\left(1+\frac{8M\chi^{4}}{3w^{3}\hbar^{2}}\right)\label{eq:m-sol}
\end{equation}
From \eqref{eq:ave-k} and \eqref{eq:m-sol}, the soliton velocity
can be predicted
\begin{equation}
v=\frac{\hbar\langle\hat{k}\rangle}{m_{\textrm{sol}}}=\frac{4rJ\left(A_{0}A_{1}+A_{1}A_{2}+A_{2}A_{3}\right)}{\hbar+\frac{128M\bar{\chi}^{4}}{3w^{3}\hbar\left(1+\xi\right)^{4}}}\sin\left(\Delta\omega\right)\label{eq:velocity}
\end{equation}

\begin{figure}[t]
\begin{centering}
\includegraphics[width=135mm]{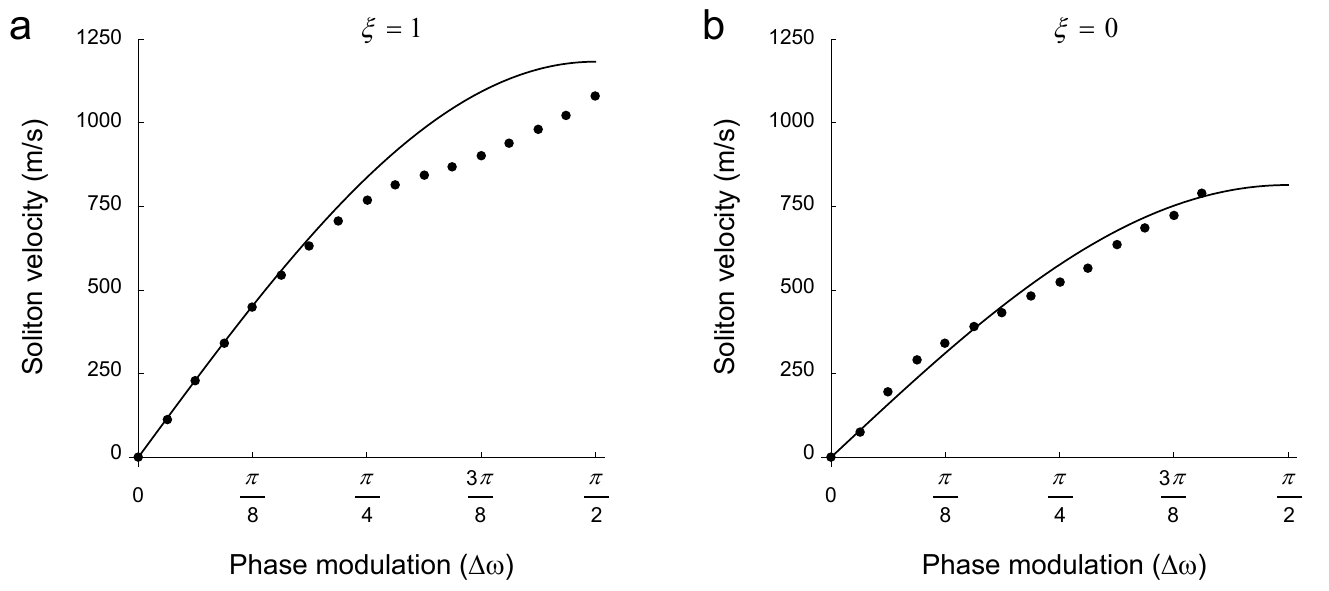}
\par\end{centering}

\caption{\label{fig:5}The velocity of single Davydov solitons for $Q=1$ plotted as a function of the phase modulation parameter $\Delta\omega$. Solitons move faster for $\xi=1$ (a) compared
with $\xi=0$ (b). The theoretical prediction of soliton velocity
based on effective soliton mass (solid line) fits well the observed
velocity (black circles) for $\xi=1$ when $\Delta\omega<\frac{\pi}{4}$
and for $\xi=0$ when $\Delta\omega<\frac{7}{16}\pi$. The generated
solitons are unstable for $\xi=0$ when $\Delta\omega\in[\frac{7}{16}\pi,\frac{\pi}{2}]$.}
\end{figure}

The velocity of single solitons ($Q=1$) observed in the simulations
was close to the predicted velocity~\eqref{eq:velocity} for $\xi=1$
when $\Delta\omega<\frac{\pi}{4}$ and for $\xi=0$ when $\Delta\omega<\frac{7}{16}\pi$
(Fig.~\ref{fig:5}). Due to larger effective mass, the solitons for
$\xi=0$ were on average 25\% slower than those for $\xi=1$. Noteworthy,
the heavier mass and lower velocity were not correlated with greater
soliton stability. Instead, the generated solitons were unstable for
$\xi=0$ when the phase modulation was $\Delta\omega\in[\frac{7}{16}\pi,\frac{\pi}{2}]$
and exhibited features of disintegration by the end of the 100 ps
simulation period (Figs. \ref{fig:3}d, \ref{fig:4}c and \ref{fig:5}b).

Collectively, the above results indicate that moving Davydov solitons
can be launched through phase-modulated Gaussian pulses of amide~I
energy even in the absence of initial phonon dressing. Thus, in the
process of evolutionary design and optimization of protein function,
it is physically plausible that different biomolecular catalysts such
as protein master proteins are able to deliver chemical energy to
protein $\alpha$-helical motifs where this energy is transported
and utilized through a soliton mechanism operating at biologically relevant
timescale of tens of picoseconds.

\subsection{Effect of phase-modulation on double solitons}

\begin{figure}[t]
\begin{centering}
\includegraphics[width=135mm]{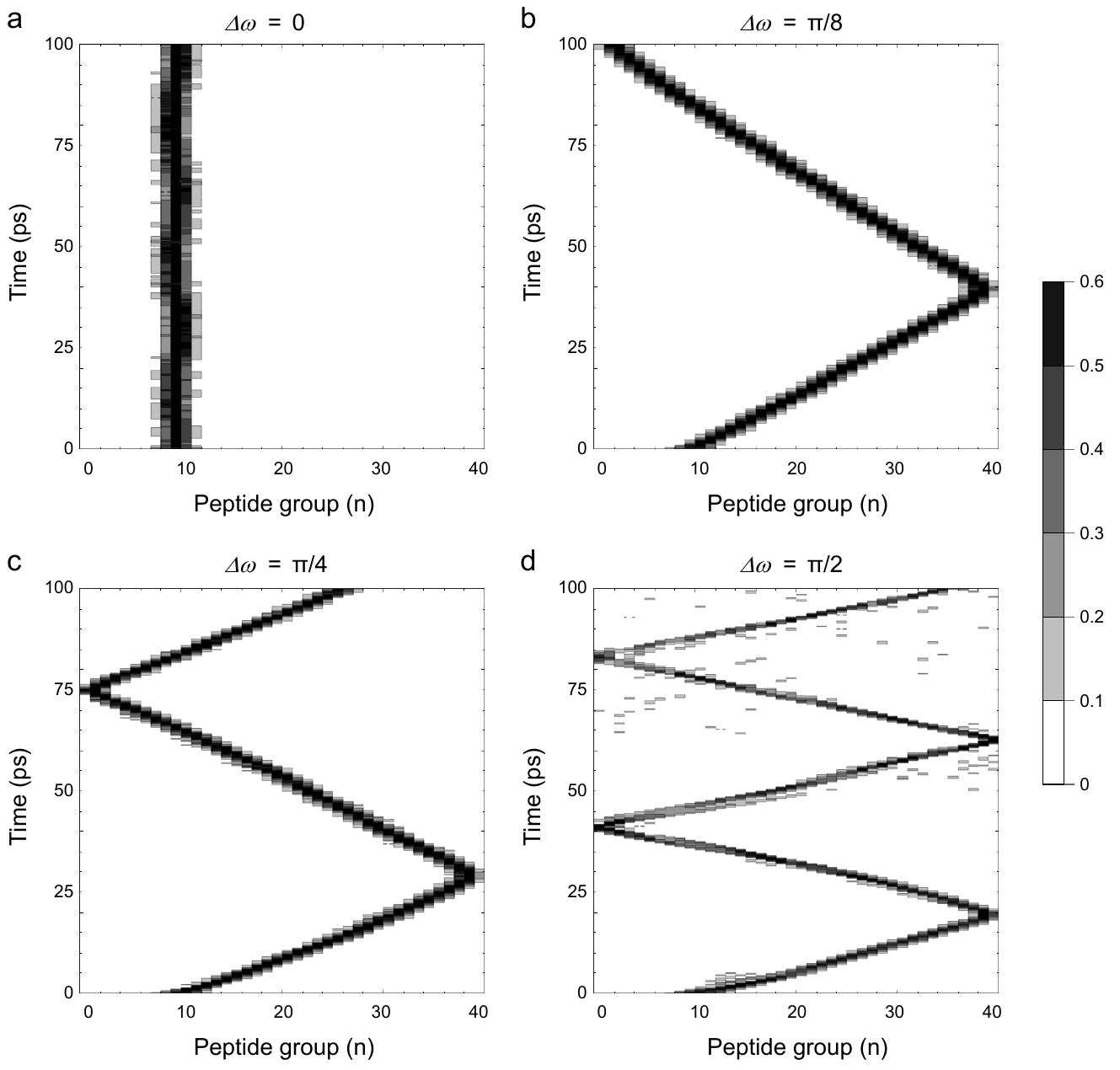}
\par\end{centering}

\caption{\label{fig:6}Double soliton dynamics for $Q=2$ and $\xi=1$, visualized in a contour plot of the amide~I exciton expectation value $Q|a_{n}|^{2}$. The applied energy pulse is centered at the peptide group $n=10$ of a protein $\alpha$-helix spine comprised of $n_{\max}=40$ peptide groups.
In the absence of phase modulation $\Delta\omega=0$ (a),
the soliton is pinned. In the presence of positive phase modulation,
$\Delta\omega=\frac{\pi}{8}$ (b), $\Delta\omega=\frac{\pi}{4}$ (c)
or $\Delta\omega=\frac{\pi}{2}$ (d), the soliton moves to the right.}
\end{figure}

\begin{figure}[t]
\begin{centering}
\includegraphics[width=135mm]{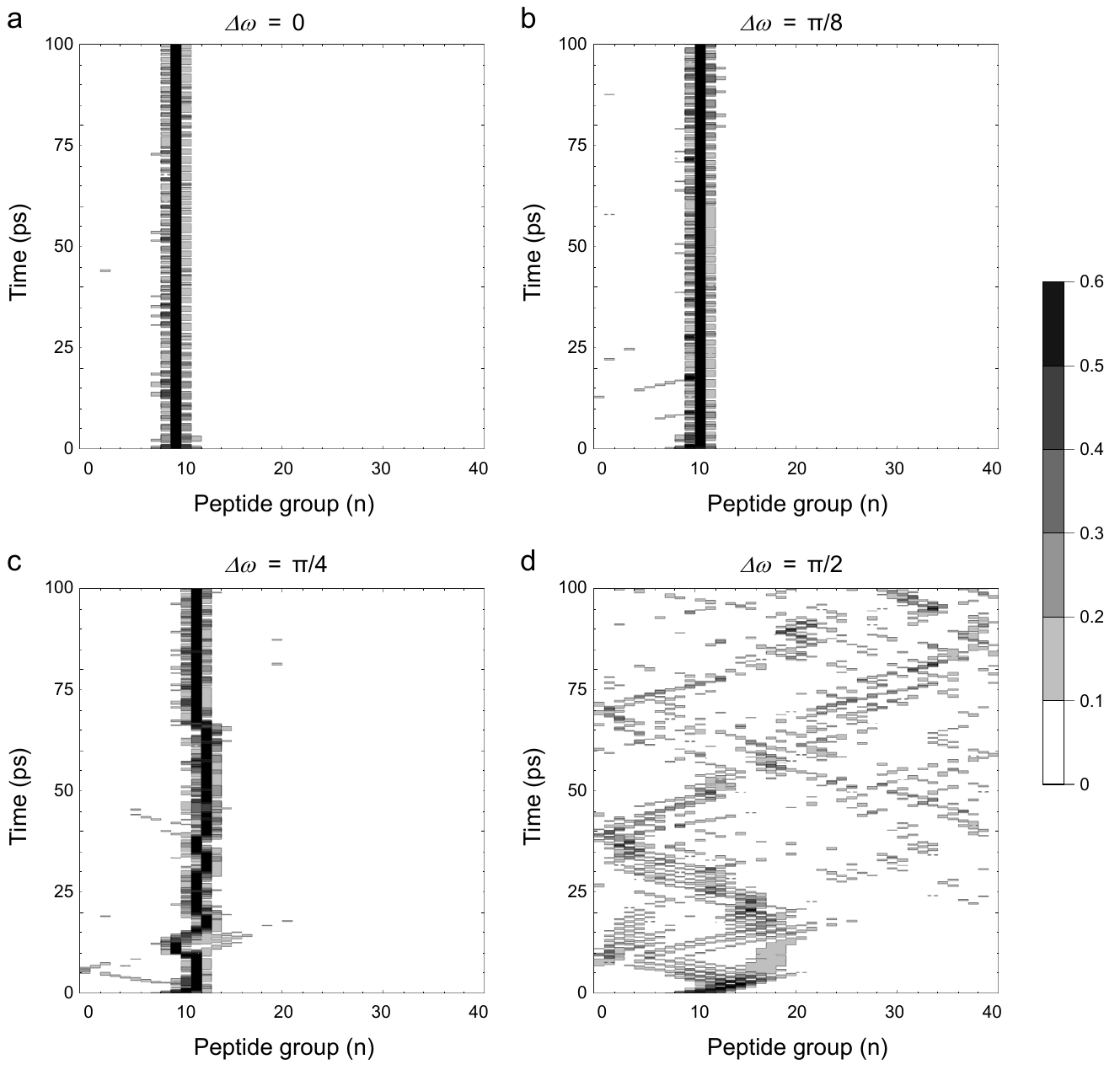}
\par\end{centering}

\caption{\label{fig:7}Double soliton dynamics for $Q=2$ and $\xi=0$, visualized in a contour plot of the amide~I exciton expectation value $Q|a_{n}|^{2}$. The applied energy pulse is centered at the peptide group $n=10$ of a protein $\alpha$-helix spine comprised of $n_{\max}=40$ peptide groups.
The soliton remains pinned despite any phase modulation, $\Delta\omega=0$
(a), $\Delta\omega=\frac{\pi}{8}$ (b) or $\Delta\omega=\frac{\pi}{4}$
(c). When the phase modulation reaches $\Delta\omega=\frac{\pi}{2}$
(d), the soliton disintegrates instead of being set in motion.}
\end{figure}

\begin{figure}[t]
\begin{centering}
\includegraphics[width=135mm]{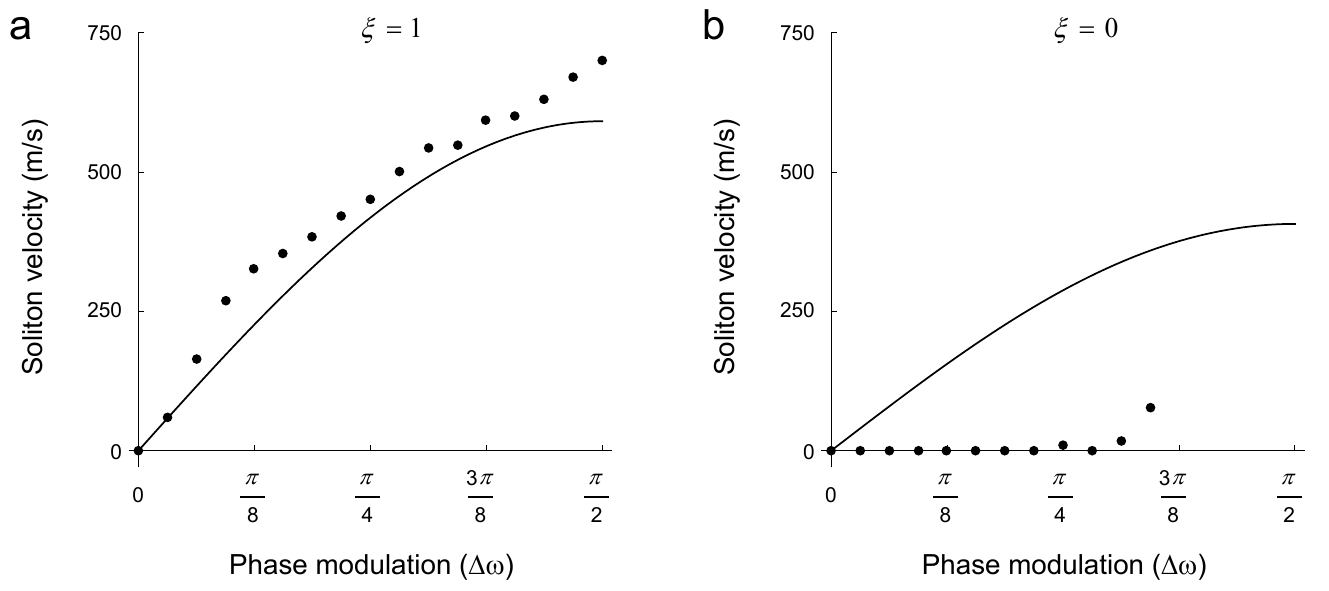}
\par\end{centering}

\caption{\label{fig:8}The velocity of double Davydov solitons for $Q=2$ plotted as a function of the phase modulation parameter $\Delta\omega$. Solitons are able to move for $\xi=1$~(a) as opposed to being pinned for $\xi=0$~(b). The theoretical prediction
of soliton velocity based on double effective soliton mass (solid
line) is close to the observed velocity (black circles) for $\xi=1$
but it is inadequate to capture the soliton pinning for $\xi=0$.
The generated solitons are unstable for $\xi=0$ when $\Delta\omega\in[\frac{7}{16}\pi,\frac{\pi}{2}]$.}
\end{figure}

To further investigate the resulting quantum dynamics in the presence
of two amide~I quanta ($Q=2$), we have generated double Davydov solitons
for different values of the phase-modulation parameter $\Delta\omega\in[0,\frac{\pi}{2}]$.
For completely isotropic exciton--phonon interaction $\xi=1$, the
non-zero phase modulation of the initial amide~I energy Gaussian distribution
was able to launch moving solitons in the interior of the protein
$\alpha$-helix (Fig.~\ref{fig:6}b-d). For completely anisotropic
exciton--phonon interaction $\xi=0$, however, the solitons remained
pinned despite the presence of significant phase modulation $\Delta\omega\in[0,\frac{3}{8}\pi)$
(Fig.~\ref{fig:7}b-c). In the range $\Delta\omega\in[\frac{\pi}{4},\frac{3}{8}\pi)$,
the pinned solitons wobbled around their place of origin, whereas
in the range $\Delta\omega\in[\frac{3}{8}\pi,\frac{\pi}{2}]$ the
solitons were unstable and disintegrated. The theoretical prediction
of soliton velocity based on double effective soliton mass was quantitatively
close to the observed velocity for $\xi=1$ (Fig.~\ref{fig:8}a),
but it was inadequate to capture the qualitative transition towards
soliton pinning for $\xi=0$ (Fig.~\ref{fig:8}b).

\begin{figure}[t]
\begin{centering}
\includegraphics[width=135mm]{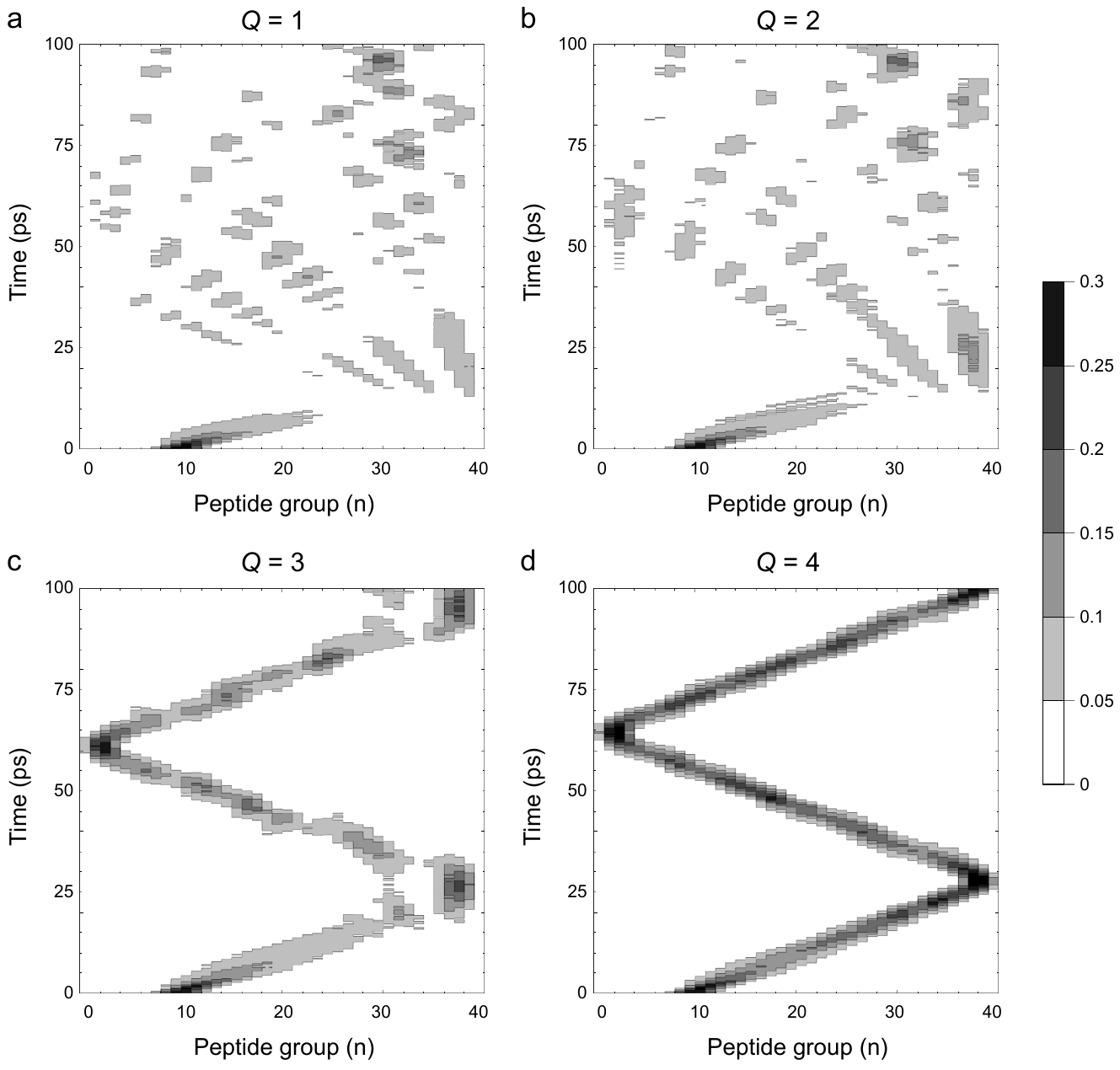}
\par\end{centering}

\caption{\label{fig:9}Launching of a phase-modulated $\Delta\omega=\frac{\pi}{8}$
Davydov soliton for $\xi=1$ in the interior of a protein $\alpha$-helix
with low value of the nonlinear exciton--phonon coupling parameter
$\bar{\chi}=15$ pN, visualized through $|a_{n}|^{2}$. Increasing
the number of amide~I excitons to $Q\geq3$ is able to strengthen
the self-trapping mechanism thereby shifting the threshold of $\bar{\chi}$
for soliton formation: $Q=1$ (a), $Q=2$ (b), $Q=3$ (c) and $Q=4$
(d).}
\end{figure}

Again collectively, the above results indicate that theoretically derived
analytic results for the soliton velocity based on effective soliton
mass might be applicable only for a narrow window of the physical
parameters in which moving soliton solutions exist. Extensive previous
computational research with the same generalized Davydov Hamiltonian
$\hat{H}=\hat{H}_\textrm{ex}+\hat{H}_\textrm{ph}+\hat{H}_\textrm{int}$
\cite{GeorgievGlazebrook2019a,GeorgievGlazebrook2019b,GeorgievGlazebrook2019c,GeorgievGlazebrook2020}
has established the existence of two different thresholds for the
nonlinear coupling parameter $\chi$: a lower threshold for which
the self-trapping mechanism is strong enough to prevent the soliton
from dispersing and an upper threshold for which the self-trapping
is so strong that the soliton remains pinned at the place of its origin
\cite{GeorgievGlazebrook2019a}. Because the number of amide~I quanta $Q$
appears as a multiplicative factor in front of $\chi$ in the
second Davydov equation \eqref{eq:gauge-2}, its main effect is to
strengthen the self-trapping mechanism thereby shifting the threshold
for soliton formation or pinning towards lower values of $\chi$.
As a result, protein structures with low nonlinear exciton--phonon
coupling might be also conductive for solitons at the expense of increasing
the number of amide~I energy quanta (Fig.~\ref{fig:9}). This would be particularly important at early stages of protein function optimization, namely a novel protein function could be initially supported by high energy flux even in the case of low nonlinear exciton--phonon coupling and then gradually the efficiency could be increased by decreasing the energy flux as the soliton stability is assisted by strengthened nonlinear exciton--phonon coupling.

\subsection{Two-soliton collisions}

\begin{figure}[t]
\begin{centering}
\includegraphics[width=135mm]{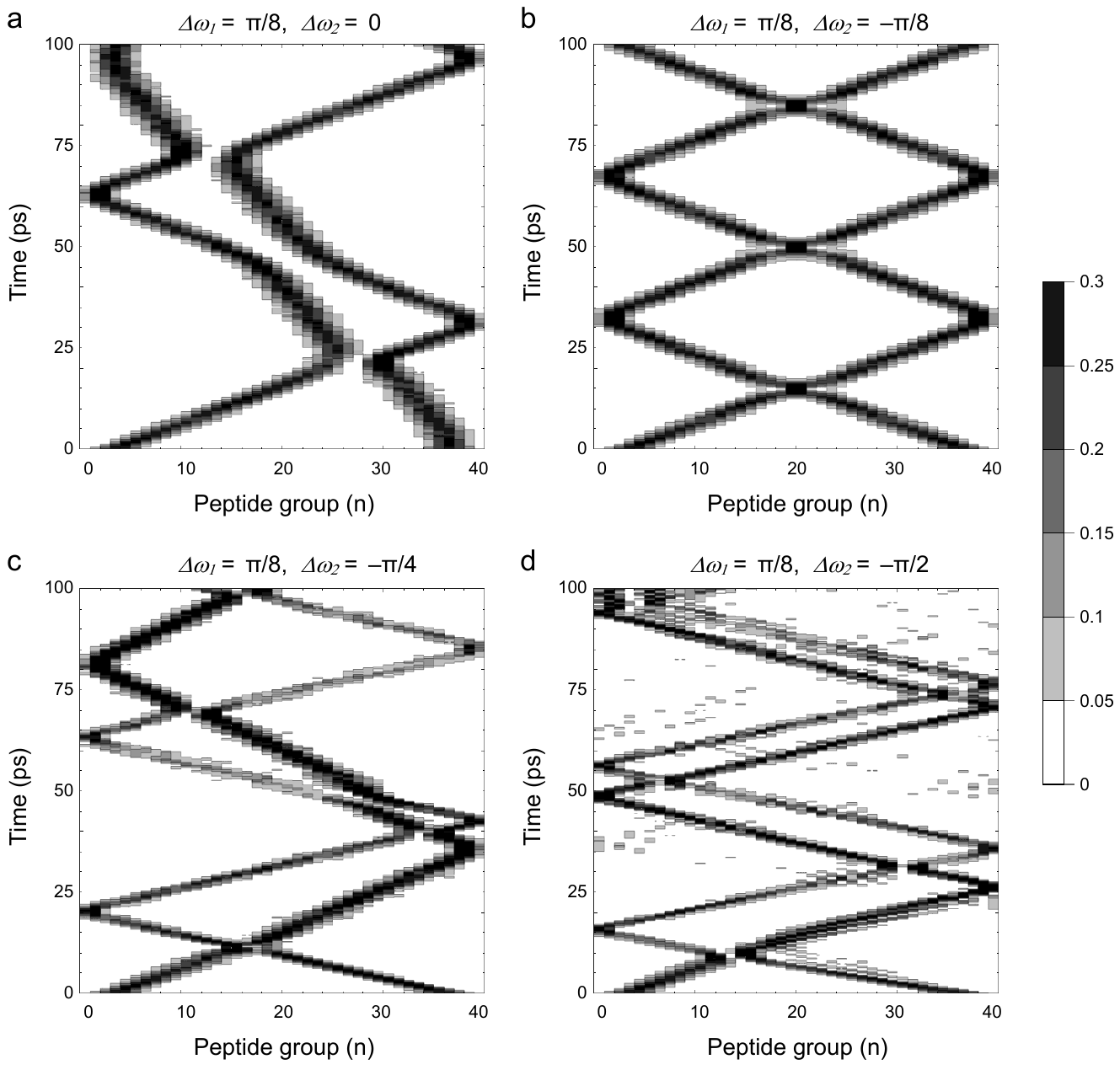}
\par\end{centering}

\caption{\label{fig:10}The quantum collision of two phase-modulated solitons
for $\xi=1$ launched from the two ends of a protein $\alpha$-helix spine,
carrying in total $Q=2$ amide~I excitons, visualized in a contour plot of $Q|a_{n}|^{2}$.
Constructive or destructive quantum interference may occur at the
collision sites. The phase modulation parameter of the soliton moving
to the right is fixed to $\Delta\omega_{1}=\frac{\pi}{8}$. The phase
modulation parameter of the soliton moving to the left varies across
different panels: $\Delta\omega_{2}=0$ (a), $\Delta\omega_{2}=-\frac{\pi}{8}$
(b), $\Delta\omega_{2}=-\frac{\pi}{4}$ (c) and $\Delta\omega_{2}=-\frac{\pi}{2}$
(d).}
\end{figure}

\begin{figure}[t]
\begin{centering}
\includegraphics[width=135mm]{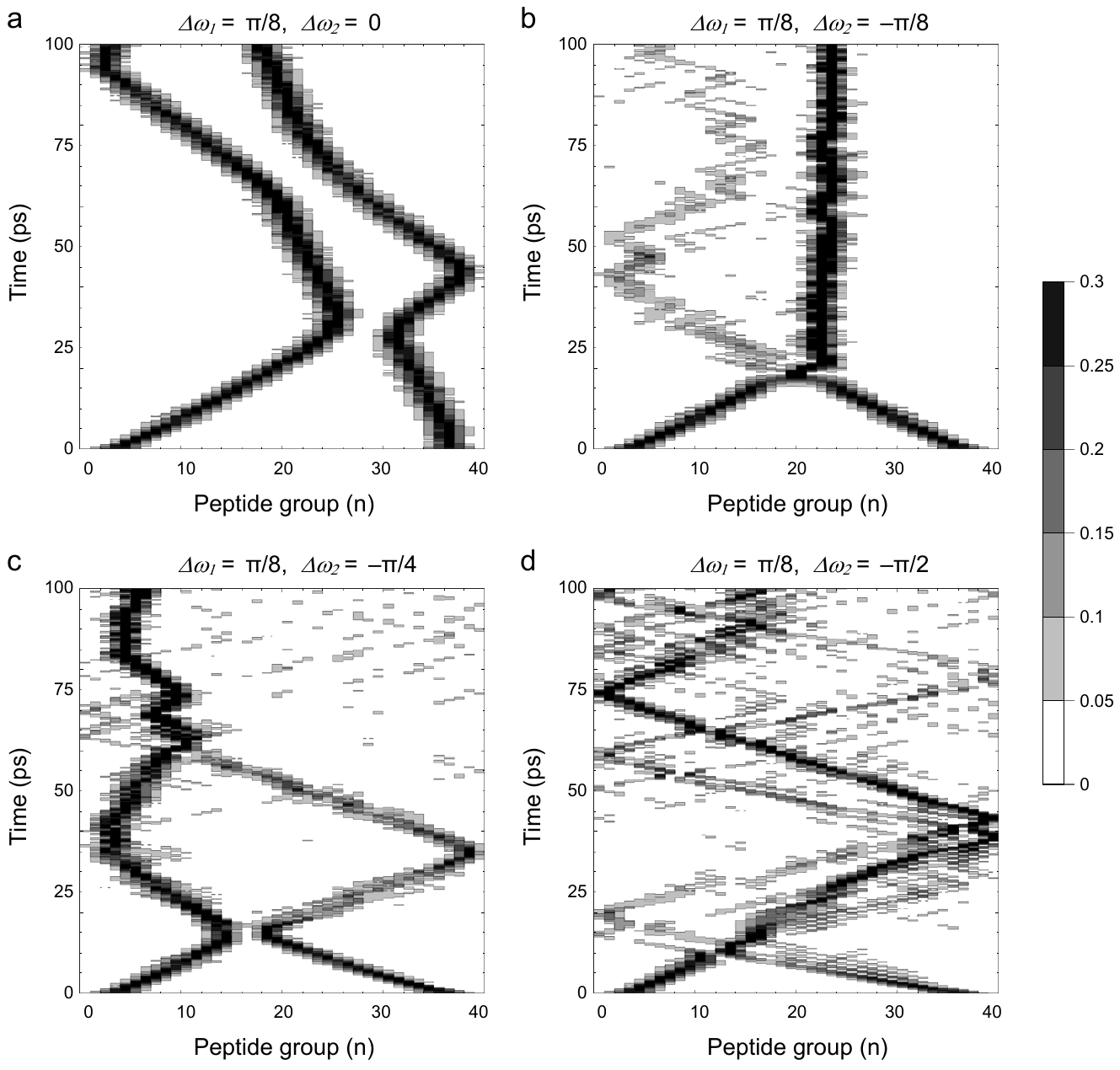}
\par\end{centering}

\caption{\label{fig:11}The quantum collision of two phase-modulated solitons
for $\xi=0$ launched from the two ends of a protein $\alpha$-helix spine,
carrying in total $Q=2$ amide~I excitons, visualized in a contour plot of $Q|a_{n}|^{2}$.
Constructive or destructive quantum interference may occur at the
collision sites. The phase modulation parameter of the soliton moving
to the right is fixed to $\Delta\omega_{1}=\frac{\pi}{8}$. The phase
modulation parameter of the soliton moving to the left varies across
different panels: $\Delta\omega_{2}=0$ (a), $\Delta\omega_{2}=-\frac{\pi}{8}$
(b), $\Delta\omega_{2}=-\frac{\pi}{4}$ (c) and $\Delta\omega_{2}=-\frac{\pi}{2}$
(d).}
\end{figure}

Having demonstrated that the phase modulation is a robust physical
mechanism that could launch moving single solitons in the interior
of the $\alpha$-helix, we have also explored how the generated solitons
interact when they collide with each other. For simulating soliton
collisions, we have used two quanta of amide~I energy ($Q=2$) and
launched two phase-modulated single solitons from the two ends of
the $\alpha$-helix thereby ensuring maximal distance between the
starting soliton positions. The soliton launched from the left
was used as a probe with fixed phase-modulation $\Delta\omega_{1}=\frac{\pi}{8}$.
The other soliton launched from the right
was then tested for different values of phase modulation $\Delta\omega_{2}\in\left\{ 0,-\frac{\pi}{8},-\frac{\pi}{4},-\frac{\pi}{2}\right\} $.
Because the negative sign of $\Delta\omega_{2}$ indicates only the
reversed direction of motion, whereas the absolute value $\left|\Delta\omega_{2}\right|$
affects the soliton speed and stability, we will report the dynamic
consequences of increasing the latter.

For $\xi=1$, the solitons exhibited constructive or destructive quantum
interference at the collision sites exemplifying the quantum nature
of the Davydov quasiparticles (Fig.~\ref{fig:10}). The colliding
solitons were also able to preserve their shape and velocity after
the collisions provided that the phase modulation of the second soliton
is in the range $\left|\Delta\omega_{2}\right|\in[0,\frac{\pi}{4}]$
(Fig.~\ref{fig:10}a-c). When the phase modulation of the second
soliton is larger, $\left|\Delta\omega_{2}\right|=\frac{\pi}{2}$,
the discreteness of the protein lattice leads to much more pronounced
leakage of exciton quantum probability amplitudes, which tends to
equalize the velocities of the two solitons recovered after each collision
(Fig.~\ref{fig:10}d).

For $\xi=0$, the solitons were more massive and moved with slower
velocities (Fig.~\ref{fig:11}). In the case of destructive quantum
interference, e.g. when $\left|\Delta\omega_{2}\right|=0$, the solitons
passed through each other preserving their original shape and velocity
(Fig.~\ref{fig:11}a). The occurrence of constructive quantum interference
at the collision sites, however, lead to formation of pinned solitons,
e.g. when $\left|\Delta\omega_{2}\right|\in\left\{ \frac{\pi}{8},\frac{\pi}{4}\right\} $
(Fig.~\ref{fig:11}b-c). Increasing the phase modulation of the second
soliton to $\left|\Delta\omega_{2}\right|=\frac{\pi}{2}$ exhibited
the instability already observed in the single soliton case, which
was further accelerated by the soliton collisions (Fig.~\ref{fig:11}d).

Taken in their entirety, the above results support the possibility that quantum
interference could have important functional consequences within living
systems \cite{GeorgievGlazebrook2020}. Indeed, constructive quantum
interference might play an important role in focusing transported energy for its utilization at protein active sites. Alternatively, destructive
quantum interference might be useful to avoid accidental absorption
of transported energy at unfavorable protein sites. Local modification
of the isotropy of exciton--phonon interaction~$\xi$ through structural
changes of the physical distances between peptide groups of sequential
amino acids may also optimize certain protein active sites to retain
the delivered energy in place through pinned solitons \cite{GeorgievGlazebrook2020}.

\section{Discussion}

Proteins sustain essential life mechanisms through catalysis of biochemical
processes in living organisms. Because the protein function involves
physical work, it can only be performed at the expense of free energy
released by biochemical reactions such as ATP hydrolysis. To investigate
deeper the quantum transport of energy inside protein $\alpha$-helices,
we have explored the generalized Davydov model that describes each $\alpha$-helix spine
as a lattice of hydrogen bonded peptide groups.
We have performed numerical simulations of the quantum dynamics of a protein
$\alpha$-helix spine for Gaussian pulses of amide~I energy applied over a region of 7 peptide groups and have confirmed the generation of either moving or stationary Davydov solitons depending on the degree of phase modulation of the initial pulses.

Previous analytical work by Davydov and colleagues \cite{Davydov1973,Davydov1976,Davydov1979,Davydov1981,Davydov1982,Davydov1986,Davydov1986b,Davydov1987,Davydov1988,Brizhik1983,Brizhik1988,Brizhik1993,Brizhik1995,Brizhik2004,Brizhik2006,Brizhik2010,Brizhik2015,Brizhik2019}
has been successful in deriving important results in regard to the
optimal shape of the solitons, as well as their energy, effective
mass and velocity. The analytic method (presented in details in \ref{sec:CA}), however, relies
upon a number of essential assumptions, satisfaction of which may be questionably
determinable in the historical early stages of life evolution. For example,
it is unlikely that proteins would have been able to deliver energy
pulses tailored precisely to the form of a sech-squared shape, or to supply an
exact initial phonon dressing for soliton production. Similarly, elaborate
mechanisms for control of soliton direction and speed may have been
lacking in prebiotic systems. Furthermore, protein $\alpha$-helical
secondary structure is sensitive to environmental conditions including
pH, and local hydrophobicity of the environment \cite{Takei2015,Ng2016,Fezoua2019},
which would impact the strength of the nonlinear exciton--phonon coupling
$\chi$, as well as the efficacy of the soliton self-trapping mechanism. In
addition, the existence of a threshold for soliton formation appears
to suggest that the transport of energy by solitons is an all-or-none
phenomenon for which natural selection is challenged to provide gradual
steps of improvement in the evolutionary history. Here, we have been
able to address all of these problems in a physically rigorous manner
supported through extensive computational simulations.

Firstly, we have established that initial phonon dressing is not absolutely
required, and applied amide~I energy pulses could have Gaussian shape
supporting the generated solitary waves for a lifetime of tens of
picoseconds. This timescale is biologically relevant because the solitons
are able to traverse the whole extent of an average protein $\alpha$-helix
and deliver the transported energy for use at a specific protein active
site. Thus, the analytic sech-squared soliton solution is to be considered
as an idealized mathematical entity whose possibly infinite lifetime
does not need to be attained. Instead, biological functionality could
be achieved by proteins with energy pulses whose Gaussian shape suffices
to support the self-trapping mechanism for only a finite lifetime
of tens of picoseconds.

Secondly, we have shown that moving solitons can be launched at any
site in the protein through phase modulation of the applied amide
I energy Gaussian pulse. Phase modulation should occur naturally since
the ATP hydrolysis sites would have different distances to different
locations in nearby protein $\alpha$-helices and quantum paths with
different lengths would accumulate different phases according to Feynman's
path integral formalism \cite{Feynman1948,Feynman1965,Kashiwa1997,Georgiev2018}.
Therefore, the relative orientation and distance between an ATP hydrolysis
site and a receptive protein $\alpha$-helix, should be sufficient
for proteins to be able to control the direction and velocity of generated
solitons.

Thirdly, we have demonstrated that low values of the nonlinear exciton--phonon
coupling~$\chi$, that are otherwise below the threshold for soliton
formation could be compensated by an increased number of amide~I quanta
in the energy pulses. Thus, in environments that destabilize the $\alpha$-helical
secondary structure, protein functionality through soliton transport
could have been achieved at the expense of higher energy fluxes. At
later evolutionary stages, when the living systems have a better control
of the relevant protein environment, thereby increasing~$\chi$, the same functionality would
be attained with fewer amide~I energy quanta.

Fourthly, we have brought into focus the existence of transient solutions
that may look like solitary waves, only for a finite period of time
after which the amide~I exciton quantum probability amplitudes disperse.
Referring to such transient solutions as solitons is in line with
Davydov's own statement that ``for describing real systems even unstable
solitary waves can be significant if their lifetime is long in comparison
with the time during which the phenomenon under study takes place''
\cite{Davydov1986b}. For Davydov, the notion of soliton (that is, as a quasiparticle)
had to be understood in a wider meaning as describing ``any autolocalized excitations propagating
without significant change in their form and velocity owing to the dynamical
balance between nonlinearity and dispersion'' \cite{Davydov1986b}.
In this general context, the most important dynamic quantity becomes
the soliton lifetime, which can be operationally defined as the time
for which the localized exciton wave loses e.g.~half of its initial
amplitude. Because the soliton lifetime has a continuous range of values,
it is subject to gradual improvement through natural selection as
mandated by evolutionary theory. This explains how proteins might have indeed evolved mechanisms for transport of energy through Davydov solitons.

Although the presented results illustrate the rich variety of physical
phenomena contained in the Davydov model, they are inevitably constrained
by our choice of biophysical parameters and the decision to focus
on the dynamics of a single protein $\alpha$-helix spine.
Quantum chemical \emph{ab initio} computation of the protein material properties utilizing recent developments in density functional theory (DFT) \cite{Kolev2013,Kolev2018} is certainly desirable as it may help revise the range of the biophysical parameters appearing in the Davydov model.
Extension of the model to a full protein $\alpha$-helix with three spines \cite{Hennig2002,Brizhik2019}
or inclusion of environmental interactions such as the presence of
external electromagnetic fields \cite{Brizhik2015} may further
inform our understanding of how proteins function.

One promising line for future research would be to perform all-atom molecular dynamics (MD) simulations, which could demonstrate soliton emergence and propagation in real $\alpha$-helical proteins. Use of atomically detailed force fields, in which interatomic interactions are considered explicitly \cite{Lindahl2015,Lopes2015}, would allow distortion of a selected C=O bond followed by running a number of protein simulations to gather sufficient statistics on the dynamics of hydrogen bonds supporting the $\alpha$-helix structure.

Another promising line of research is to use a generalized discrete nonlinear Schr\"{o}dinger equation (DNLSE) arising from a C$_\alpha$-trace-based energy function to study topological solitons involved in protein folding \cite{Khalili2005a,Khalili2005b,Chernodub2010,Molkenthin2011}. Using their united-residue (UNRES) force field simulations, with respect to the staphylococcal protein A,
Niemi and coworkers already did substantial work on soliton propagation and energy transport in proteins at a coarse-grained level \cite{Krokhotin2012a,Krokhotin2012b,Krokhotin2014}. Although the topological solitons there are manifested as perturbation of geometric chain geometry-symmetry, and not hydrogen-bond network (involving C=O distortions), the coarse-grained description is expected to be intimately related to the asymmetric three-spine Davydov soliton, which spontaneously breaks the local translational and helical symmetries \cite{Brizhik2004}. In particular, asymmetric stretching, or contraction of the hydrogen bonds in the three parallel $\alpha$-helix spines will induce a topological kink in the protein, and conversely, the presence of a topological kink will inevitably distort the hydrogen bonds in the protein $\alpha$-helix spines. The free energy of kink formation in the coarse-grained molecular dynamics was observed to be $\approx 7$ kcal/mol, which roughly corresponds to a double Davydov soliton with $Q=2$ amide~I quanta. Because the Davydov model is also mathematically based on DNLSE, our findings in regard to the effects of phase modulation on soliton velocity might be also applicable to protein folding, and thus we are not too far apart from the results reported in Ref.~\cite{Krokhotin2014}. For example, in-phase pulses of energy are expected to produce stationary topological kinks in the protein structure, whereas phase-modulated pulses of energy are expected to produce traveling kinks that transport energy, and hence subserve the functionally important motions of proteins.

\section*{Conflict of interest}

The authors declare that they have no conflict of interest.

\appendix

\section{Continuum approximation}
\label{sec:CA}

The continuum approximation employs the mathematical transformations
$f_{n}\to f\left(x,t\right)$ and $f_{n\pm1}\to\left[1\pm r\frac{\partial}{\partial x}+\frac{r^{2}}{2}\frac{\partial^{2}}{\partial x^{2}}\right]f\left(x,t\right)$
in order to replace the set of discrete functions in the system of
generalized Davydov equations \eqref{eq:gauge-1} and \eqref{eq:gauge-2}
with corresponding continuous functions. The system of discrete ordinary differential equations (ODEs),
then becomes a system of two partial differential equations (PDEs) in terms of the exciton distribution
$a\left(x,t\right)$ and the phonon displacements $b\left(x,t\right)$.
For ease of notation, we will leave the spatial and temporal dependence
of $a\left(x,t\right)$ and $b\left(x,t\right)$ implicit
\begin{eqnarray}
\imath\hbar\frac{\partial a}{\partial t} & = & -2Ja-Jr^{2}\frac{\partial^{2}a}{\partial x^{2}}+\chi a\left[\left(1+\xi\right)r\frac{\partial b}{\partial x}+\left(1-\xi\right)\frac{r^{2}}{2}\frac{\partial^{2}b}{\partial x^{2}}\right]\label{eq:PDE-1}\\
M\frac{\partial^{2}b}{\partial t^{2}} & = & wr^{2}\frac{\partial^{2}b}{\partial x^{2}}+Q\chi\left[\left(1+\xi\right)r\frac{\partial\left|a\right|^{2}}{\partial x}-\left(1-\xi\right)\frac{r^{2}}{2}\frac{\partial^{2}\left|a\right|^{2}}{\partial x^{2}}\right]\label{eq:PDE-2}
\end{eqnarray}
Searching for a solution traveling at a constant speed $v$ such that
$\frac{\partial^{2}}{\partial t^{2}}b=v^{2}\frac{\partial^{2}}{\partial x^{2}}b$,
we can rearrange \eqref{eq:PDE-2} in the form
\begin{equation}
\frac{\partial^{2}b}{\partial x^{2}}=-\frac{Q\chi}{wr}\frac{1}{\left(1-s^{2}\right)}\left[\left(1+\xi\right)\frac{\partial\left|a\right|^{2}}{\partial x}-\left(1-\xi\right)\frac{r}{2}\frac{\partial^{2}\left|a\right|^{2}}{\partial x^{2}}\right]\label{eq:Dav-1}
\end{equation}
where $s=\frac{v}{v_{0}}$ and $v_{0}=r\sqrt{\frac{w}{M}}$ is the
velocity of longitudinal sound waves in the chain of peptide groups
\cite{Davydov1979,Davydov1986b}. Integrating once with respect to
$x$ and setting the integration constant to zero gives
\begin{equation}
\frac{\partial b}{\partial x}=-\frac{Q\chi}{wr}\frac{1}{\left(1-s^{2}\right)}\left[\left(1+\xi\right)\left|a\right|^{2}-\left(1-\xi\right)\frac{r}{2}\frac{\partial\left|a\right|^{2}}{\partial x}\right]\label{eq:Dav-2}
\end{equation}
After substitution of \eqref{eq:Dav-1} and \eqref{eq:Dav-2} in \eqref{eq:PDE-1},
we obtain
\begin{equation}
\imath\hbar\frac{\partial a}{\partial t}=-2Ja-Jr^{2}\frac{\partial^{2}a}{\partial x^{2}}-a\frac{Q\chi^{2}}{w\left(1-s^{2}\right)}\left[\left(1+\xi\right)^{2}\left|a\right|^{2}-\left(1-\xi\right)^{2}\frac{r^{2}}{4}\frac{\partial^{2}\left|a\right|^{2}}{\partial x^{2}}\right]
\end{equation}
If we consider the isotropic case $\xi=1$, which was originally proposed
by Davydov \cite{Davydov1979}, we obtain the nonlinear Schr\"{o}dinger
equation (NLSE) \cite{Taghizadeh2011,Taghizadeh2017} in a standard form
\begin{equation}
\left[\imath\frac{\partial}{\partial t}+\alpha+\beta\frac{\partial^{2}}{\partial x^{2}}+\gamma\left|a\right|^{2}\right]a=0\label{eq:NLSE}
\end{equation}
with $\alpha=\frac{2J}{\hbar}$, $\beta=\frac{Jr^{2}}{\hbar}$ and
$\gamma=\frac{4Q\chi^{2}}{\hbar w\left(1-s^{2}\right)}$. The analytic
soliton solution centered at $x_{0}$ is
\begin{equation}
a\left(x,t\right)=\sqrt{\frac{2}{\gamma}}\textrm{sech}\left[\sqrt{\frac{1}{\beta}}\left(x-x_{0}-vt\right)\right]\exp\left[\imath\left(\alpha+1\right)t+\imath\frac{v}{2\beta}\left(x-x_{0}-\frac{1}{2}vt\right)\right]
\end{equation}
with corresponding sech-squared exciton probability distribution
\begin{equation}
\left|a\left(x,t\right)\right|^{2}=\frac{2}{\gamma}\textrm{sech}^{2}\left[\sqrt{\frac{1}{\beta}}\left(x-x_{0}-vt\right)\right]
\end{equation}

\section{Supplementary videos}

\paragraph{Video~1}
Quantum dynamics of a Davydov soliton carrying a single amide~I exciton ($Q=1$), simulated for 100 ps with completely isotropic exciton--phonon interaction $\xi=1$, and visualized through $\textrm{Re}\left(a_{n}\right)$ and $\textrm{Im}\left(a_{n}\right)$ of the exciton quantum probability amplitudes. The applied amide~I energy pulse is centered at the peptide group $n=10$ of a protein $\alpha$-helix spine comprised of $n_{\max}=40$ peptide groups.
In the absence of phase modulation $\Delta\omega=0$, the soliton is pinned.

\paragraph{Video~2}
The quantum dynamics of a Davydov soliton carrying a single amide~I exciton ($Q=1$), simulated for 100 ps with completely isotropic exciton--phonon interaction $\xi=1$, and visualized through $\textrm{Re}\left(a_{n}\right)$ and $\textrm{Im}\left(a_{n}\right)$ of the exciton quantum probability amplitudes. The applied amide~I energy pulse is centered at the peptide group $n=10$ of a protein $\alpha$-helix spine comprised of $n_{\max}=40$ peptide groups.
In the presence of phase modulation $\Delta\omega=\frac{\pi}{4}$, the soliton moves
to the right.

\paragraph{Video~3}
The quantum dynamics of a Davydov soliton carrying a single amide~I exciton ($Q=1$), simulated for 100 ps with completely isotropic exciton--phonon interaction $\xi=1$, and visualized through $\textrm{Re}\left(a_{n}\right)$ and $\textrm{Im}\left(a_{n}\right)$ of the exciton quantum probability amplitudes. The applied amide~I energy pulse is centered at the peptide group $n=10$ of a protein $\alpha$-helix spine comprised of $n_{\max}=40$ peptide groups.
In the presence of phase modulation $\Delta\omega=\frac{\pi}{2}$, the soliton moves
faster to the right but starts to lose amplitude.

\paragraph{Video~4}
The quantum dynamics of a Davydov soliton carrying a single amide~I exciton ($Q=1$), simulated for 100 ps with completely isotropic exciton--phonon interaction $\xi=1$, and visualized through $\textrm{Re}\left(a_{n}\right)$ and $\textrm{Im}\left(a_{n}\right)$ of the exciton quantum probability amplitudes. The applied amide~I energy pulse is centered at the peptide group $n=10$ of a protein $\alpha$-helix spine comprised of $n_{\max}=40$ peptide groups.
In the presence of phase modulation $\Delta\omega=\pi$, the soliton is unstable and quickly disintegrates.


\begin{thebibliography}{10}
\expandafter\ifx\csname url\endcsname\relax
  \def\url#1{\texttt{#1}}\fi
\expandafter\ifx\csname urlprefix\endcsname\relax\def\urlprefix{URL }\fi
\expandafter\ifx\csname href\endcsname\relax
  \def\href#1#2{#2} \def\path#1{#1}\fi

\bibitem{GeorgievGlazebrook2020}
D.~D. Georgiev, J.~F. Glazebrook, Quantum transport and utilization of free
  energy in protein $\alpha$-helices, Advances in Quantum Chemistry 82 (2020)
  in press.
\newblock \href {https://doi.org/10.1016/bs.aiq.2020.02.001}
  {\path{doi:10.1016/bs.aiq.2020.02.001}}.

\bibitem{Scholtz1992}
J.~M. Scholtz, R.~L. Baldwin, The mechanism of $\alpha$-helix formation by
  peptides, Annual Review of Biophysics and Biomolecular Structure 21~(1)
  (1992) 95--118.
\newblock \href {https://doi.org/10.1146/annurev.bb.21.060192.000523}
  {\path{doi:10.1146/annurev.bb.21.060192.000523}}.

\bibitem{Kohn1997}
W.~D. Kohn, C.~T. Mant, R.~S. Hodges, $\alpha$-helical protein assembly motifs,
  Journal of Biological Chemistry 272~(5) (1997) 2583--2586.
\newblock \href {https://doi.org/10.1074/jbc.272.5.2583}
  {\path{doi:10.1074/jbc.272.5.2583}}.

\bibitem{Haimov2016}
B.~Haimov, S.~Srebnik, A closer look into the $\alpha$-helix basin, Scientific
  Reports 6~(1) (2016) 38341.
\newblock \href {https://doi.org/10.1038/srep38341}
  {\path{doi:10.1038/srep38341}}.

\bibitem{Pauling1951a}
L.~Pauling, R.~B. Corey, H.~R. Branson, The structure of proteins: two
  hydrogen-bonded helical configurations of the polypeptide chain, Proceedings
  of the National Academy of Sciences 37~(4) (1951) 205--211.
\newblock \href {https://doi.org/10.1073/pnas.37.4.205}
  {\path{doi:10.1073/pnas.37.4.205}}.

\bibitem{GeorgievGlazebrook2019a}
D.~D. Georgiev, J.~F. Glazebrook, On the quantum dynamics of {D}avydov solitons
  in protein $\alpha$-helices, Physica A: Statistical Mechanics and its
  Applications 517 (2019) 257--269.
\newblock \href {https://doi.org/10.1016/j.physa.2018.11.026}
  {\path{doi:10.1016/j.physa.2018.11.026}}.

\bibitem{Davydov1973}
A.~S. Davydov, The theory of contraction of proteins under their excitation,
  Journal of Theoretical Biology 38~(3) (1973) 559--569.
\newblock \href {https://doi.org/10.1016/0022-5193(73)90256-7}
  {\path{doi:10.1016/0022-5193(73)90256-7}}.

\bibitem{Davydov1976}
A.~S. Davydov, N.~I. Kislukha, Solitons in one-dimensional molecular chains,
  Physica Status Solidi (b) 75~(2) (1976) 735--742.
\newblock \href {https://doi.org/10.1002/pssb.2220750238}
  {\path{doi:10.1002/pssb.2220750238}}.

\bibitem{Davydov1979}
A.~S. Davydov, Solitons in molecular systems, Physica Scripta 20~(3-4) (1979)
  387--394.
\newblock \href {https://doi.org/10.1088/0031-8949/20/3-4/013}
  {\path{doi:10.1088/0031-8949/20/3-4/013}}.

\bibitem{Davydov1981}
A.~S. Davydov, The role of solitons in the energy and electron transfer in
  one-dimensional molecular systems, Physica D 3~(1-2) (1981) 1--22.
\newblock \href {https://doi.org/10.1016/0167-2789(81)90116-0}
  {\path{doi:10.1016/0167-2789(81)90116-0}}.

\bibitem{Davydov1982}
A.~S. Davydov, Solitons in quasi-one-dimensional molecular structures, Soviet
  Physics Uspekhi 25~(12) (1982) 898--918.
\newblock \href {https://doi.org/10.1070/pu1982v025n12abeh005012}
  {\path{doi:10.1070/pu1982v025n12abeh005012}}.

\bibitem{Davydov1986}
A.~S. Davydov, Quantum theory of the motion of a quasi-particle in a molecular
  chain with thermal vibrations taken into account, Physica Status Solidi (b)
  138~(2) (1986) 559--576.
\newblock \href {https://doi.org/10.1002/pssb.2221380221}
  {\path{doi:10.1002/pssb.2221380221}}.

\bibitem{Davydov1986b}
A.~S. Davydov, Solitons in biology, in: S.~E. Trullinger, V.~E. Zakharov, V.~L.
  Pokrovsky (Eds.), Solitons, Modern Problems in Condensed Matter Sciences,
  North-Holland, Amsterdam, 1986, pp. 1--51.

\bibitem{Davydov1987}
A.~S. Davydov, V.~N. Ermakov, Linear and nonlinear resonance electron tunneling
  through a system of potential barriers, Physica D: Nonlinear Phenomena 28~(1)
  (1987) 168--180.
\newblock \href {https://doi.org/10.1016/0167-2789(87)90127-8}
  {\path{doi:10.1016/0167-2789(87)90127-8}}.

\bibitem{Davydov1988}
A.~S. Davydov, V.~N. Ermakov, Soliton generation at the boundary of a molecular
  chain, Physica D: Nonlinear Phenomena 32~(2) (1988) 318--323.
\newblock \href {https://doi.org/10.1016/0167-2789(88)90059-0}
  {\path{doi:10.1016/0167-2789(88)90059-0}}.

\bibitem{Kivshar1989}
Y.~S. Kivshar, B.~A. Malomed, Dynamics of solitons in nearly integrable
  systems, Reviews of Modern Physics 61~(4) (1989) 763--915.
\newblock \href {https://doi.org/10.1103/RevModPhys.61.763}
  {\path{doi:10.1103/RevModPhys.61.763}}.

\bibitem{Brizhik1983}
L.~S. Brizhik, A.~S. Davydov, Soliton excitations in one-dimensional molecular
  systems, Physica Status Solidi (b) 115~(2) (1983) 615--630.
\newblock \href {https://doi.org/10.1002/pssb.2221150233}
  {\path{doi:10.1002/pssb.2221150233}}.

\bibitem{Brizhik1988}
L.~S. Brizhik, Y.~B. Gaididei, A.~A. Vakhnenko, V.~A. Vakhnenko, Soliton
  generation in semi-infinite molecular chains, Physica Status Solidi (b)
  146~(2) (1988) 605--612.
\newblock \href {https://doi.org/10.1002/pssb.2221460221}
  {\path{doi:10.1002/pssb.2221460221}}.

\bibitem{Brizhik1993}
L.~S. Brizhik, Soliton generation in molecular chains, Physical Review B 48~(5)
  (1993) 3142--3144.
\newblock \href {https://doi.org/10.1103/PhysRevB.48.3142}
  {\path{doi:10.1103/PhysRevB.48.3142}}.

\bibitem{Brizhik1995}
L.~S. Brizhik, A.~A. Eremko, Electron autolocalized states in molecular chains,
  Physica D: Nonlinear Phenomena 81~(3) (1995) 295--304.
\newblock \href {https://doi.org/10.1016/0167-2789(94)00206-6}
  {\path{doi:10.1016/0167-2789(94)00206-6}}.

\bibitem{Brizhik2004}
L.~S. Brizhik, A.~Eremko, B.~Piette, W.~Zakrzewski, Solitons in
  $\alpha$-helical proteins, Physical Review E 70~(3) (2004) 031914.
\newblock \href {https://doi.org/10.1103/PhysRevE.70.031914}
  {\path{doi:10.1103/PhysRevE.70.031914}}.

\bibitem{Brizhik2006}
L.~Brizhik, A.~Eremko, B.~Piette, W.~Zakrzewski, Charge and energy transfer by
  solitons in low-dimensional nanosystems with helical structure, Chemical
  Physics 324~(1) (2006) 259--266.
\newblock \href {https://doi.org/10.1016/j.chemphys.2006.01.033}
  {\path{doi:10.1016/j.chemphys.2006.01.033}}.

\bibitem{Brizhik2010}
L.~Brizhik, A.~Eremko, B.~Piette, W.~Zakrzewski, Ratchet effect of {D}avydov's
  solitons in nonlinear low-dimensional nanosystems, International Journal of
  Quantum Chemistry 110~(1) (2010) 25--37.
\newblock \href {https://doi.org/10.1002/qua.22083}
  {\path{doi:10.1002/qua.22083}}.

\bibitem{Cruzeiro1988}
L.~Cruzeiro, J.~Halding, P.~L. Christiansen, O.~Skovgaard, A.~C. Scott,
  Temperature effects on the {D}avydov soliton, Physical Review A 37~(3) (1988)
  880--887.
\newblock \href {https://doi.org/10.1103/PhysRevA.37.880}
  {\path{doi:10.1103/PhysRevA.37.880}}.

\bibitem{Cruzeiro1994}
L.~Cruzeiro-Hansson, V.~M. Kenkre, Localized versus delocalized ground states
  of the semiclassical {H}olstein {H}amiltonian, Physics Letters A 190~(1)
  (1994) 59--64.
\newblock \href {https://doi.org/10.1016/0375-9601(94)90366-2}
  {\path{doi:10.1016/0375-9601(94)90366-2}}.

\bibitem{Cruzeiro1997}
L.~Cruzeiro-Hansson, S.~Takeno, Davydov model: the quantum, mixed
  quantum-classical, and full classical systems, Physical Review E 56~(1)
  (1997) 894--906.
\newblock \href {https://doi.org/10.1103/PhysRevE.56.894}
  {\path{doi:10.1103/PhysRevE.56.894}}.

\bibitem{Cruzeiro2009}
L.~Cruzeiro, The {D}avydov/{S}cott model for energy storage and transport in
  proteins, Journal of Biological Physics 35~(1) (2009) 43--55.
\newblock \href {https://doi.org/10.1007/s10867-009-9129-0}
  {\path{doi:10.1007/s10867-009-9129-0}}.

\bibitem{Kerr1987}
W.~C. Kerr, P.~S. Lomdahl, Quantum-mechanical derivation of the equations of
  motion for {D}avydov solitons, Physical Review B 35~(7) (1987) 3629--3632.
\newblock \href {https://doi.org/10.1103/PhysRevB.35.3629}
  {\path{doi:10.1103/PhysRevB.35.3629}}.

\bibitem{Kerr1990}
W.~C. Kerr, P.~S. Lomdahl, Quantum-mechanical derivation of the {D}avydov
  equations for multi-quanta states, in: P.~L. Christiansen, A.~C. Scott
  (Eds.), Davydov's Soliton Revisited: Self-Trapping of Vibrational Energy in
  Protein, Springer, New York, 1990, pp. 23--30.
\newblock \href {https://doi.org/10.1007/978-1-4757-9948-4_2}
  {\path{doi:10.1007/978-1-4757-9948-4_2}}.

\bibitem{MacNeil1984}
L.~MacNeil, A.~C. Scott, Launching a {D}avydov soliton: {II}. {N}umerical
  studies, Physica Scripta 29~(3) (1984) 284--287.
\newblock \href {https://doi.org/10.1088/0031-8949/29/3/017}
  {\path{doi:10.1088/0031-8949/29/3/017}}.

\bibitem{Scott1984}
A.~C. Scott, Launching a {D}avydov soliton: {I}. {S}oliton analysis, Physica
  Scripta 29~(3) (1984) 279--283.
\newblock \href {https://doi.org/10.1088/0031-8949/29/3/016}
  {\path{doi:10.1088/0031-8949/29/3/016}}.

\bibitem{Scott1985}
A.~C. Scott, Davydov solitons in polypeptides, Philosophical Transactions of
  the Royal Society of London Series A, Mathematical and Physical Sciences
  315~(1533) (1985) 423--436.
\newblock \href {https://doi.org/10.1098/rsta.1985.0049}
  {\path{doi:10.1098/rsta.1985.0049}}.

\bibitem{Scott1992}
A.~C. Scott, Davydov's soliton, Physics Reports 217~(1) (1992) 1--67.
\newblock \href {https://doi.org/10.1016/0370-1573(92)90093-F}
  {\path{doi:10.1016/0370-1573(92)90093-F}}.

\bibitem{Luo2011}
B.~Luo, J.~Ye, Y.~Zhao, Variational study of polaron dynamics with the
  {D}avydov ans\"{a}tze, Physica Status Solidi (c) 8~(1) (2011) 70--73.
\newblock \href {https://doi.org/10.1002/pssc.201000721}
  {\path{doi:10.1002/pssc.201000721}}.

\bibitem{Luo2017}
J.~Luo, B.~Piette, A generalised {D}avydov-{S}cott model for polarons in linear
  peptide chains, European Physical Journal B 90~(8) (2017) 155.
\newblock \href {https://doi.org/10.1140/epjb/e2017-80209-2}
  {\path{doi:10.1140/epjb/e2017-80209-2}}.

\bibitem{Sun2010}
J.~Sun, B.~Luo, Y.~Zhao, Dynamics of a one-dimensional {H}olstein polaron with
  the {D}avydov ans\"{a}tze, Physical Review B 82~(1) (2010) 014305.
\newblock \href {https://doi.org/10.1103/PhysRevB.82.014305}
  {\path{doi:10.1103/PhysRevB.82.014305}}.

\bibitem{GeorgievGlazebrook2019b}
D.~D. Georgiev, J.~F. Glazebrook, Quantum tunneling of {D}avydov solitons
  through massive barriers, Chaos, Solitons and Fractals 123 (2019) 275--293.
\newblock \href {https://doi.org/10.1016/j.chaos.2019.04.013}
  {\path{doi:10.1016/j.chaos.2019.04.013}}.

\bibitem{Petzold1983}
L.~Petzold, Automatic selection of methods for solving stiff and nonstiff
  systems of ordinary differential equations, SIAM Journal on Scientific and
  Statistical Computing 4~(1) (1983) 136--148.
\newblock \href {https://doi.org/10.1137/0904010} {\path{doi:10.1137/0904010}}.

\bibitem{Hindmarsh1983}
A.~C. Hindmarsh, {ODEPACK}, a systematized collection of {ODE} solvers, in:
  R.~S. Stepleman, M.~Carver, R.~Peskin, W.~F. Ames, R.~Vichnevetsky (Eds.),
  Scientific Computing, {IMACS} Transactions on Scientific Computation,
  North-Holland, Amsterdam, 1983, pp. 55--64.

\bibitem{Hindmarsh1995}
A.~C. Hindmarsh, L.~R. Petzold, Algorithms and software for ordinary
  differential equations and differential-algebraic equations, {P}art {II}:
  {H}igher-order methods and software packages, Computers in Physics 9~(2)
  (1995) 148--155.
\newblock \href {https://doi.org/10.1063/1.168540}
  {\path{doi:10.1063/1.168540}}.

\bibitem{Trott2006}
M.~Trott, The Mathematica GuideBook for Numerics, Springer, New York, 2006.
\newblock \href {https://doi.org/10.1007/0-387-28814-7}
  {\path{doi:10.1007/0-387-28814-7}}.

\bibitem{Brizhik2015}
L.~S. Brizhik, Influence of electromagnetic field on soliton-mediated charge
  transport in biological systems, Electromagnetic Biology and Medicine 34~(2)
  (2015) 123--132.
\newblock \href {https://doi.org/10.3109/15368378.2015.1036071}
  {\path{doi:10.3109/15368378.2015.1036071}}.

\bibitem{Itoh1972}
K.~Itoh, T.~Shimanouchi, Vibrational spectra of crystalline formamide, Journal
  of Molecular Spectroscopy 42~(1) (1972) 86--99.
\newblock \href {https://doi.org/10.1016/0022-2852(72)90146-4}
  {\path{doi:10.1016/0022-2852(72)90146-4}}.

\bibitem{Nevskaya1976}
N.~A. Nevskaya, Y.~N. Chirgadze, Infrared spectra and resonance interactions of
  amide-{I} and {II} vibrations of $\alpha$-helix, Biopolymers 15~(4) (1976)
  637--648.
\newblock \href {https://doi.org/10.1002/bip.1976.360150404}
  {\path{doi:10.1002/bip.1976.360150404}}.

\bibitem{Lai1989a}
Y.~Lai, H.~A. Haus, Quantum theory of solitons in optical fibers. {I}.
  {T}ime-dependent {H}artree approximation, Physical Review A 40~(2) (1989)
  844--853.
\newblock \href {https://doi.org/10.1103/PhysRevA.40.844}
  {\path{doi:10.1103/PhysRevA.40.844}}.

\bibitem{Lai1989b}
Y.~Lai, H.~A. Haus, Quantum theory of solitons in optical fibers. {II}. {E}xact
  solution, Physical Review A 40~(2) (1989) 854--866.
\newblock \href {https://doi.org/10.1103/PhysRevA.40.854}
  {\path{doi:10.1103/PhysRevA.40.854}}.

\bibitem{Wright1990}
E.~M. Wright, Quantum theory of self-phase modulation, Journal of the Optical
  Society of America B 7~(6) (1990) 1142--1146.
\newblock \href {https://doi.org/10.1364/josab.7.001142}
  {\path{doi:10.1364/josab.7.001142}}.

\bibitem{Glauber1963a}
R.~J. Glauber, The quantum theory of optical coherence, Physical Review 130~(6)
  (1963) 2529--2539.
\newblock \href {https://doi.org/10.1103/PhysRev.130.2529}
  {\path{doi:10.1103/PhysRev.130.2529}}.

\bibitem{Glauber1963b}
R.~J. Glauber, Coherent and incoherent states of the radiation field, Physical
  Review 131~(6) (1963) 2766--2788.
\newblock \href {https://doi.org/10.1103/PhysRev.131.2766}
  {\path{doi:10.1103/PhysRev.131.2766}}.

\bibitem{GeorgievGlazebrook2019c}
D.~D. Georgiev, J.~F. Glazebrook, Neurotransmitter release and conformational
  changes within the {SNARE} protein complex, in: S.~E. Lyshevski (Ed.),
  Nanoengineering, Quantum Science, and, Nanotechnology Handbook, CRC Press,
  Boca Raton, 2019, pp. 375--404.

\bibitem{Brizhik2019}
L.~S. Brizhik, J.~Luo, B.~M. A.~G. Piette, W.~J. Zakrzewski, Long-range
  donor-acceptor electron transport mediated by $\alpha$-helices, Physical
  Review E 100~(6) (2019) 062205.
\newblock \href {https://doi.org/10.1103/PhysRevE.100.062205}
  {\path{doi:10.1103/PhysRevE.100.062205}}.

\bibitem{Takei2015}
T.~Takei, K.~Tsumoto, A.~Okonogi, A.~Kimura, S.~Kojima, K.~Yazaki, T.~Takei,
  T.~Ueda, K.~Miura, p{H} responsiveness of fibrous assemblies of
  repeat-sequence amphipathic $\alpha$-helix polypeptides, Protein Science
  24~(5) (2015) 883--894.
\newblock \href {https://doi.org/10.1002/pro.2665}
  {\path{doi:10.1002/pro.2665}}.

\bibitem{Ng2016}
D.~P. Ng, C.~M. Deber, Modulating transmembrane $\alpha$-helix interactions
  through p{H}-sensitive boundary residues, Biochemistry 55~(31) (2016)
  4306--4315.
\newblock \href {https://doi.org/10.1021/acs.biochem.6b00380}
  {\path{doi:10.1021/acs.biochem.6b00380}}.

\bibitem{Fezoua2019}
Z.~Fezoua-Boubegtiten, B.~Hastoy, P.~Scotti, A.~Milochau, K.~Bathany,
  B.~Desbat, S.~Castano, R.~Oda, J.~Lang, The transmembrane domain of the
  {SNARE} protein {VAMP2} is highly sensitive to its lipid environment,
  Biochimica et Biophysica Acta (BBA) - Biomembranes 1861~(3) (2019) 670--676.
\newblock \href {https://doi.org/10.1016/j.bbamem.2018.12.011}
  {\path{doi:10.1016/j.bbamem.2018.12.011}}.

\bibitem{Feynman1948}
R.~P. Feynman, Space-time approach to non-relativistic quantum mechanics,
  Reviews of Modern Physics 20~(2) (1948) 367--387.
\newblock \href {https://doi.org/10.1103/RevModPhys.20.367}
  {\path{doi:10.1103/RevModPhys.20.367}}.

\bibitem{Feynman1965}
R.~P. Feynman, A.~R. Hibbs, Quantum Mechanics and Path Integrals, McGraw-Hill,
  New York, 1965.

\bibitem{Kashiwa1997}
T.~Kashiwa, Y.~Ohnuki, M.~Suzuki, Path Integral Methods, Oxford University
  Press, Oxford, 1997.

\bibitem{Georgiev2018}
D.~D. Georgiev, E.~Cohen, Probing finite coarse-grained virtual {F}eynman
  histories with sequential weak values, Physical Review A 97~(5) (2018)
  052102.
\newblock \href {https://doi.org/10.1103/PhysRevA.97.052102}
  {\path{doi:10.1103/PhysRevA.97.052102}}.

\bibitem{Kolev2013}
S.~K. Kolev, P.~S. Petkov, M.~Rangelov, G.~N. Vayssilov, Ab initio molecular
  dynamics of {N}a$^+$ and {M}g$^{2+}$ countercations at the backbone of {RNA}
  in water solution, ACS Chemical Biology 8~(7) (2013) 1576--1589.
\newblock \href {https://doi.org/10.1021/cb300463h}
  {\path{doi:10.1021/cb300463h}}.

\bibitem{Kolev2018}
S.~K. Kolev, P.~S. Petkov, M.~A. Rangelov, D.~V. Trifonov, T.~I. Milenov, G.~N.
  Vayssilov, Interaction of {N}a$^+$, {K}$^+$, {M}g$^{2+}$ and {C}a$^{2+}$
  counter cations with {RNA}, Metallomics 10~(5) (2018) 659--678.
\newblock \href {https://doi.org/10.1039/c8mt00043c}
  {\path{doi:10.1039/c8mt00043c}}.

\bibitem{Hennig2002}
D.~Hennig, Energy transport in $\alpha$-helical protein models: one-strand
  versus three-strand systems, Physical Review B 65~(17) (2002) 174302.
\newblock \href {https://doi.org/10.1103/PhysRevB.65.174302}
  {\path{doi:10.1103/PhysRevB.65.174302}}.

\bibitem{Lindahl2015}
E.~Lindahl, Molecular dynamics simulations, in: A.~Kukol (Ed.), Molecular
  Modeling of Proteins, Vol. 1215 of Methods in Molecular Biology, Springer,
  New York, 2015, pp. 3--26.
\newblock \href {https://doi.org/10.1007/978-1-4939-1465-4_1}
  {\path{doi:10.1007/978-1-4939-1465-4_1}}.

\bibitem{Lopes2015}
P.~E.~M. Lopes, O.~Guvench, A.~D. MacKerell, Current status of protein force
  fields for molecular dynamics simulations, in: A.~Kukol (Ed.), Molecular
  Modeling of Proteins, Vol. 1215 of Methods in Molecular Biology, Springer,
  New York, 2015, pp. 47--71.
\newblock \href {https://doi.org/10.1007/978-1-4939-1465-4_3}
  {\path{doi:10.1007/978-1-4939-1465-4_3}}.

\bibitem{Khalili2005a}
M.~Khalili, A.~Liwo, F.~Rakowski, P.~Grochowski, H.~A. Scheraga, Molecular
  dynamics with the united-residue model of polypeptide chains. {I}. {L}agrange
  equations of motion and tests of numerical stability in the microcanonical
  mode, Journal of Physical Chemistry B 109~(28) (2005) 13785--13797.
\newblock \href {https://doi.org/10.1021/jp058008o}
  {\path{doi:10.1021/jp058008o}}.

\bibitem{Khalili2005b}
M.~Khalili, A.~Liwo, A.~Jagielska, H.~A. Scheraga, Molecular dynamics with the
  united-residue model of polypeptide chains. {II}. {L}angevin and
  {B}erendsen-bath dynamics and tests on model $\alpha$-helical systems,
  Journal of Physical Chemistry B 109~(28) (2005) 13798--13810.
\newblock \href {https://doi.org/10.1021/jp058007w}
  {\path{doi:10.1021/jp058007w}}.

\bibitem{Chernodub2010}
M.~Chernodub, S.~Hu, A.~J. Niemi, Topological solitons and folded proteins,
  Physical Review E 82~(1) (2010) 011916.
\newblock \href {https://doi.org/10.1103/PhysRevE.82.011916}
  {\path{doi:10.1103/PhysRevE.82.011916}}.

\bibitem{Molkenthin2011}
N.~Molkenthin, S.~Hu, A.~J. Niemi, Discrete nonlinear {S}chr\"{o}dinger equation
  and polygonal solitons with applications to collapsed proteins, Physical
  Review Letters 106~(7) (2011) 078102.
\newblock \href {https://doi.org/10.1103/PhysRevLett.106.078102}
  {\path{doi:10.1103/PhysRevLett.106.078102}}.

\bibitem{Krokhotin2012a}
A.~Krokhotin, A.~J. Niemi, X.~Peng, Soliton concepts and protein structure,
  Physical Review E 85~(3) (2012) 031906.
\newblock \href {https://doi.org/10.1103/PhysRevE.85.031906}
  {\path{doi:10.1103/PhysRevE.85.031906}}.

\bibitem{Krokhotin2012b}
A.~Krokhotin, M.~Lundgren, A.~J. Niemi, Solitons and collapse in the
  $\lambda$-repressor protein, Physical Review E 86~(2) (2012) 021923.
\newblock \href {https://doi.org/10.1103/PhysRevE.86.021923}
  {\path{doi:10.1103/PhysRevE.86.021923}}.

\bibitem{Krokhotin2014}
A.~Krokhotin, A.~Liwo, G.~G. Maisuradze, A.~J. Niemi, H.~A. Scheraga, Kinks,
  loops, and protein folding, with protein {A} as an example, Journal of
  Chemical Physics 140~(2) (2014) 025101.
\newblock \href {https://doi.org/10.1063/1.4855735}
  {\path{doi:10.1063/1.4855735}}.

\bibitem{Taghizadeh2011}
N.~Taghizadeh, M.~Mirzazadeh, F.~Farahrooz, Exact solutions of the nonlinear
  {S}chr\"{o}dinger equation by the first integral method, Journal of
  Mathematical Analysis and Applications 374~(2) (2011) 549--553.
\newblock \href {https://doi.org/10.1016/j.jmaa.2010.08.050}
  {\path{doi:10.1016/j.jmaa.2010.08.050}}.

\bibitem{Taghizadeh2017}
N.~Taghizadeh, Q.~Zhou, M.~Ekici, M.~Mirzazadeh, Soliton solutions for
  {D}avydov solitons in $\alpha$-helix proteins, Superlattices and
  Microstructures 102 (2017) 323--341.
\newblock \href {https://doi.org/10.1016/j.spmi.2016.12.057}
  {\path{doi:10.1016/j.spmi.2016.12.057}}.

\end{thebibliography}
\end{document}